\documentclass[aps,prd,preprint,groupedaddress,nofootinbib,titlepage]{revtex4-2}

\usepackage{amsmath,amssymb,amsthm,amsxtra,overpic,bbm,bm,epsfig,subfigure}
\usepackage{hyperref}
\usepackage{mathrsfs}
\usepackage{enumitem}
\usepackage{graphicx}
\usepackage{color}
\usepackage[dvipsnames]{xcolor}
\usepackage{comment}
\usepackage{epstopdf}
\usepackage{float}

\usepackage{slashed,stmaryrd}
\usepackage{youngtab}
\usepackage{makecell}

\usepackage{diagbox}
\usepackage{blkarray}

\allowdisplaybreaks[1]
\hypersetup{
	colorlinks =true,
	linkcolor  = red,
	citecolor  = blue,
	urlcolor   = blue
}

\newcommand{\Op}{\mathcal{O}}
\newcommand{\Opr}{\mathcal{R}}
\newcommand{\loopf}{\frac{1}{16\pi^2 \varepsilon}}

\newcommand{\rmi}{{\rm i}}

\newcommand{\Dls}{\mbox{$\raisebox{2mm}{\boldmath ${}^\leftarrow$}\hspace{-4mm}\slashed{D}$}}

\begin{document}

\preprint{TUM-HEP 1512/24}


\title{Threshold Effects on the Massless Neutrino in the Canonical Seesaw Mechanism}

\author{Di Zhang}
\email[]{di1.zhang@tum.de}
\affiliation{Physik-Department, Technische Universität München, James-Franck-Straße, 85748 Garching, Germany}

\begin{abstract}
	In this work, we revisit the one-loop renormalization group equations (RGEs) among non-degenerate seesaw scales, i.e., threshold effects in the canonical seesaw mechanism, which have been obtained for more than two decades. Different from the previous work only focusing on the Weinberg operator, we derive the complete one-loop RGEs of all three dimension-five operators in the Standard Model effective field theory with right-handed neutrinos ($\nu$SMEFT) and apply them to threshold effects in the canonical seesaw mechanism. We find some contributions from the Weinberg operator to its Wilson coefficient, the neutrino Yukawa coupling matrix, and the Higgs quartic coupling absent in the previous calculations. Based on the updated one-loop RGEs, we derive the RGE of the effective neutrino mass matrix's determinant without any approximation. Then, for the first time, we provide a strict proof that the one-loop RG running effects among non-degenerate seesaw scales can not generate a non-zero mass for the initial massless neutrino in the minimal type-I seesaw mechanism or in the canonical one with a rank-degenerate neutrino Yukawa coupling matrix. One has to include two- or higher-loop corrections to achieve a non-zero mass for the massless neutrino.
\end{abstract}

\maketitle

\section{\label{sec:introduction}Introduction}
Robust evidences from neutrino oscillation experiments with different neutrino sources manifest that at least two active neutrinos have non-zero masses~\cite{ParticleDataGroup:2022pth,Xing:2020ijf}. However, the absolute neutrino mass scale is still unknown and the lightest neutrino may even be massless. The latter case can naturally show up at the tree level in the minimal type-I seesaw mechanism~\cite{Kleppe:1995zz,Ma:1998zg} (see Ref.~\cite{Xing:2020ald} for the latest review), which extends the Standard Model with two right-handed neutrinos and is a reduction of the canonical one with three right-handed neutrinos~\cite{Minkowski:1977sc,Yanagida:1979as,Gell-Mann:1979vob,Glashow:1979nm,Mohapatra:1979ia}. The canonical seesaw mechanism can also lead to a massless neutrino with a rank-two neutrino Yukawa coupling matrix. One natural and intriguing question is how stable this zero neutrino mass is against radiative corrections since no fundamental symmetry prevents it from getting a non-zero value. This fantastic issue has been discussed by taking into account fixed-scale corrections~\cite{Petcov:1984nz,Babu:1988ig,Grimus:1989pu,Aparici:2011nu,Schmidt:2011jp,Aparici:2012vx,Pilaftsis:1991ug} or renormalization-group (RG) running corrections without threshold effects~\cite{Antusch:2005gp,Davidson:2006tg,Ibarra:2018dib,Ibarra:2020eia,Bonilla:2020sne,Xing:2020ezi,Zhou:2021bqs} (see, e.g., Refs.~\cite{Cai:2017jrq} and~\cite{Ohlsson:2013xva} for a review on these two kinds of quantum corrections, respectively). It turns out that this vanishing mass in the (minimal) type-I seesaw mechanism is stable against one-loop corrections, but after two-loop corrections are introduced, the lightest neutrino can get a non-vanishing mass~\cite{Petcov:1984nz,Babu:1988ig,Grimus:1989pu,Aparici:2011nu,Schmidt:2011jp,Aparici:2012vx,Davidson:2006tg,Ibarra:2018dib,Ibarra:2020eia,Bonilla:2020sne,Xing:2020ezi}. However, for the one-loop RG effects with non-degenerate seesaw scales, it is still not clear whether the above conclusion holds when the threshold corrections among seesaw scales are taking into consideration. Recently, Ref.~\cite{Benoit:2022ohv} has attempted to explore such threshold effects on the vanishing neutrino mass, but no definite result has been achieved for the canonical seesaw mechanism.

In this work, we provide a strict proof of threshold effects on the initially vanishing neutrino mass in a different way from that in Ref.~\cite{Benoit:2022ohv}. Before that, we revisit the renormalization group equations (RGEs) among non-degenerate seesaw scales, which were derived more than two decades ago~\cite{Antusch:2002rr}. Instead of focusing on the Weinberg operator, we work in the Standard Model effective field theory with right-handed neutrinos, i.e., the so-called $\nu$SMEFT~\cite{Anisimov:2006hv,delAguila:2008ir,Anisimov:2008gg,Aparici:2009fh}, which can describe the canonical seesaw mechanism with heavier right-handed neutrino(s) integrated out. We firstly achieve the complete one-loop RGEs of all three dimension-five (dim-5) operators in the $\nu$SMEFT and then apply results to the threshold effects in the (minimal) type-I seesaw mechanism. We find that among seesaw scales, the Weinberg operator gives some contributions to the one-loop RGEs of the neutrino Yukawa coupling matrix and the Higgs quartic coupling, besides that of its own Wilson coefficient. These contributions are overlooked in the previous derivations~\cite{Antusch:2002rr} and make the discussions in Ref.~\cite{Benoit:2022ohv} invalid. It shows that the effective neutrino mass matrix is not largely affected by the new contributions due to a cancellation among them. But the neutrino Yukawa coupling matrix and the Wilson coefficient of the Weinberg operator are indeed influenced separately, especially their flavor structures. Based on those complete one-loop RGEs, we derived the differential equation of the effective neutrino mass matrix's determinant against the renormalization scale both for the minimal type-I seesaw mechanism and for the canonical one without any approximation. In the former case, the determinant keeps vanishing during the whole energy range. In the latter case, we obtain the integral form of the determinant, which is proportional to its initial value. This indicates that the determinant will remain vanishing if it initially vanishes, despite including threshold effects. Therefore, for the first time, we strictly prove that threshold effects among non-degenerate seesaw scales by the one-loop RGEs can not generate a non-zero mass for the initially massless neutrino in the minimal type-I seesaw mechanism or in the canonical one with a rank-degenerate neutrino Yukawa coupling matrix. In these two cases, a non-zero mass generated for the initially massless neutrino first appears at the two-loop level~\cite{Davidson:2006tg,Ibarra:2018dib,Ibarra:2020eia,Xing:2020ezi,Ibarra:2024in}. Some interesting RG running behaviors of the determinant are also explored. 

\section{\label{sec:nuSMEFT}RGEs in the $\nu$SMEFT up to Dimension Five}

Before discussing threshold effects in the canonical seesaw mechanism, we take into consideration a general framework, i.e., the $\nu$SMEFT~\cite{Anisimov:2006hv,delAguila:2008ir,Anisimov:2008gg,Aparici:2009fh}, which contains three physical dim-5 operators including the Weinberg operator and is applicable to threshold effects in the canonical seesaw mechanism.
The Lagrangian of the $\nu$SMEFT up to dimension five can be written as
\begin{eqnarray}\label{eq:LnuSMEFT}
	\mathcal{L}^{}_{\nu \rm SMEFT} &=& \mathcal{L}^{}_{\rm SM} + \mathcal{L}^{}_N + \mathcal{L}^{}_{\rm dim-5} \;
\end{eqnarray}
with
\begin{eqnarray}\label{eq:LSM}
	\mathcal{L}^{}_{\rm SM} &=& - \frac{1}{4} G^{A}_{\mu\nu} G^{A\mu\nu} - \frac{1}{4} W^I_{\mu\nu} W^{I\mu\nu} - \frac{1}{4} B^{}_{\mu\nu} B^{\mu\nu} + \left( D^{}_\mu H \right)^\dagger \left( D^\mu H \right) - m^2 H^\dagger H - \lambda \left( H^\dagger H \right)^2 
	\nonumber
	\\
	&& + \sum^{}_f \overline{f} \rmi \slashed{D} f - \left[ \overline{Q^{}_{\alpha \rm L}} \left(Y^{}_{\rm u} \right)^{}_{\alpha \beta} \widetilde{H} U^{}_{\beta \rm R} +  \overline{Q^{}_{\alpha \rm L}}  (Y^{}_{\rm d})^{}_{\alpha \beta} H D^{}_{\beta \rm R}+ \overline{\ell^{}_{\alpha \rm L}}(Y^{}_l)^{}_{\alpha \beta} H E^{}_{\beta \rm R}+ {\rm h.c.} \right]  \;,
	\nonumber
	\\
	\mathcal{L}^{}_{N} &=& \overline{N^{}_{\rm R}} \rmi \slashed{\partial} N^{}_{\rm R} - \left( \frac{1}{2} \overline{N^{\rm c}_{\rm R}} M^{}_N N^{}_{\rm R} + \overline{\ell^{}_{\rm L}} Y^{}_\nu \widetilde{H} N^{}_{\rm R} + {\rm h.c.} \right) \;,
	\nonumber
	\\
	\mathcal{L}^{}_{\rm dim-5} &=& \frac{1}{2} C^{\alpha\beta}_5 \Op^{\alpha\beta}_5 + C^{\alpha\beta}_{HN} \Op^{\alpha\beta}_{HN} + C^{\alpha\beta}_{BN} \Op^{\alpha\beta}_{BN} + {\rm h.c.} \;,
\end{eqnarray}
where $f= Q^{}_{\rm L}, U^{}_{\rm R}, D^{}_{\rm R}, \ell^{}_{\rm L}, E^{}_{\rm R}$ is introduced and the three dim-5 operators are~\cite{delAguila:2008ir,Aparici:2009fh}
\begin{eqnarray}\label{eq:dim5operators}
	\Op^{\alpha\beta}_5 = \overline{\ell^{}_{\alpha\rm L}} \widetilde{H} \widetilde{H}^{\rm T} \ell^{\rm c}_{\beta \rm L} \;,\quad \Op^{\alpha\beta}_{HN} = \overline{N^{\rm c}_{\alpha \rm R}} N^{}_{\beta \rm R} H^\dagger H \;,\quad \Op^{\alpha\beta}_{BN} = \overline{N^{\rm c}_{\alpha\rm R}} \rmi \sigma^{\mu\nu} N^{}_{\beta\rm R} B^{}_{\mu\nu} \;.
\end{eqnarray}
As seen from Eq.~\eqref{eq:dim5operators}, besides the Weinberg operator~\cite{Weinberg:1979sa}, there are two additional dim-5 operators involving right-handed neutrinos in the $\nu$SMEFT compared to those in the SMEFT~\cite{Buchmuller:1985jz,Grzadkowski:2010es}. Usually, the so-called Green's basis~\cite{Jiang:2018pbd} is invovled if one renormalizes the Lagrangian containing operators of higher mass dimension with the off-shell scheme. The Green's basis for dim-5 operators in the $\nu$SMEFT contains three redundant dim-5 operators in additional to the three physical ones in Eq.~\eqref{eq:dim5operators}, which are chosen as
\begin{eqnarray}\label{eq:dim5roperators}
	R^{\alpha\beta}_{DN} = \overline{N^{\rm c}_{\alpha\rm R}} \partial^2 N^{}_{\beta\rm R} \;,\quad R^{\alpha\beta}_{\ell HN1} = \overline{\ell^{}_{\alpha\rm L}} \widetilde{H} \slashed{\partial} N^{\rm c}_{\beta\rm R} \;,\quad R^{\alpha\beta}_{\ell HN2} = \overline{\ell^{}_{\alpha\rm L}}  \Dls  \widetilde{H} N^{\rm c}_{\beta\rm R} \;.
\end{eqnarray} 
These three redundant operators appear in intermediate calculations and need to be converted into physical ones by means of equations of motion (EoMs) or field redefinitions. Following the calculation strategy in Refs.~\cite{Zhang:2023kvw,Zhang:2023ndw}, one can derive one-loop RGEs for all renormalizable couplings and Wilson coefficients of the three dim-5 operators in Eq.~\eqref{eq:dim5operators} with the modified subtraction scheme. The one-loop calculations are straightforward but lengthy, which we are not going to show here~\footnote{In Appendix~\ref{app:A}, we give two examples in detail, where the operator $R^{\alpha\beta}_{\ell HN1}$ given in Eq.~\eqref{eq:dim5roperators} plays a significant role.}.  The results for the renormalizable couplings are given by
\begin{eqnarray}\label{eq:rge-coupling}
	\mu \frac{{\rm d} g^{}_1}{{\rm d} \mu} &=& \frac{41g^3_1}{96\pi^2}  \;,\qquad
	\mu \frac{{\rm d} g^{}_2}{{\rm d} \mu} = - \frac{19g^3_2}{96\pi^2} \;,\qquad
	\mu \frac{{\rm d} g^{}_3}{{\rm d} \mu} = - \frac{7g^3_3}{16\pi^2} \;,
	\nonumber
	\\
	\mu \frac{{\rm d} m^2}{{\rm d} \mu} &=& \frac{1}{16\pi^2} \left[ \left( - \frac{3}{2} g^2_1 - \frac{9}{2}g^2_2 + 12\lambda + 2T \right) m^2 - 4{\rm Tr} \left( M^\dagger_N M^{}_N Y^\dagger_\nu Y^{}_\nu \right) \right.
	\nonumber
	\\
	&&  + \left.  4{\rm Tr} \left( M^\dagger_N M^{}_N M^\dagger_N C^{}_{HN} + M^{}_N M^\dagger_N M^{}_N C^\dagger_{HN}  \right)  \right] \;,
	\nonumber
	\\
	\mu \frac{{\rm d} \lambda}{{\rm d} \mu}  &=& \frac{1}{16\pi^2 } \left[ 4\lambda T + \frac{3}{8} \left( g^2_1+g^2_2 \right)^2 + \frac{3}{4}g^4_2 - 3\lambda \left( g^2_1 + 3g^2_2 \right) + 24 \lambda^2 - 2 T^\prime \right.
	\nonumber
	\\
	&& + \left.  2{\rm Tr} \left( 4 C^\dagger_{HN} M^{}_N Y^\dagger_\nu Y^{}_\nu + 4Y^\dagger_\nu Y^{}_\nu M^\dagger_N C^{}_{HN} + C^{}_5 Y^\ast_\nu M^{}_N Y^\dagger_\nu + Y^{}_\nu M^\dagger_N Y^{\rm T}_\nu C^\dagger_5 \right) \right] \;,
	\nonumber
	\\
	\mu \frac{{\rm d} Y^{}_\nu }{{\rm d} \mu} &=&  \frac{1}{16\pi^2} \left[ Y^{}_\nu \left(T - \frac{3}{4}g^2_1 - \frac{9}{4} g^2_2 \right)  + \frac{3}{2} \left( Y^{}_\nu Y^\dagger_\nu - Y^{}_l Y^\dagger_l   \right) Y^{}_\nu + 12 \rmi g^{}_1 Y^{}_\nu M^\dagger_N C^{}_{BN} \right.
	\nonumber
	\\
	&& - \left.  4 Y^{}_\nu M^\dagger_N C^{}_{HN} - 3C^{}_5 Y^\ast_\nu M^{}_N \right] \;,
	\nonumber
	\\
	\mu \frac{{\rm d} Y^{}_l }{{\rm d} \mu} &=& \frac{1}{16\pi^2 } \left[ Y^{}_l \left( T - \frac{15}{4}g^2_1 - \frac{9}{4}g^2_2 \right)  + \frac{3}{2} \left( Y^{}_l Y^\dagger_l  -  Y^{}_\nu Y^\dagger_\nu \right) Y^{}_l\right] \;,
	\nonumber
	\\
	\mu \frac{{\rm d} Y^{}_{\rm u} }{{\rm d} \mu} &=& \frac{1}{16\pi^2 } \left[ Y^{}_{\rm u} \left( T - \frac{17}{12}g^2_1 - \frac{9}{4}g^2_2 - 8 g^2_3 \right)  + \frac{3}{2} \left( Y^{}_{\rm u} Y^\dagger_{\rm u} - Y^{}_{\rm d} Y^\dagger_{\rm d} \right) Y^{}_{\rm u}  \right] \;,
	\nonumber
	\\
	\mu \frac{{\rm d} Y^{}_{\rm d} }{{\rm d} \mu}  &=& \frac{1}{16\pi^2}  \left[ Y^{}_{\rm d} \left( T - \frac{5}{12}g^2_1 - \frac{9}{4}g^2_2 - 8 g^2_3 \right)  - \frac{3}{2} \left( Y^{}_{\rm u} Y^\dagger_{\rm u} - Y^{}_{\rm d} Y^\dagger_{\rm d}  \right) Y^{}_{\rm d}  \right] \;,
	\nonumber
	\\
	\mu \frac{{\rm d} M^{}_N}{{\rm d} \mu}&=& \frac{1}{16\pi^2} \left[ M^{}_N Y^\dagger_\nu Y^{}_\nu + \left( Y^\dagger_\nu Y^{}_\nu \right)^{\rm T} M^{}_N -  8 m^2 C^{}_{HN}  \right] \;,
\end{eqnarray}
and those for the Wilson coefficients of the three physical operators are  found to be
\begin{eqnarray}\label{eq:rge-wc}
	\mu \frac{{\rm d} C^{}_5 }{{\rm d} \mu} &=&  \frac{1}{16\pi^2}  \left[ \left( 4\lambda -3g^2_2 + 2T \right)C^{}_5  +  \left(  \frac{7}{2} Y^{}_\nu Y^\dagger_\nu - \frac{3}{2}Y^{}_l Y^\dagger_l \right) C^{}_5 +  C^{}_5 \left( \frac{7}{2} Y^{}_\nu Y^\dagger_\nu - \frac{3}{2} Y^{}_lY^\dagger_l  \right)^{\rm T} \right] \;,
	\nonumber
	\\
	\mu \frac{{\rm d} C^{}_{BN}}{{\rm d} \mu} &=& \frac{1}{16\pi^2}  \left[ \frac{41}{6}g^2_1 C^{}_{BN} +  C^{}_{BN}Y^\dagger_\nu Y^{}_\nu + \left( Y^\dagger_\nu Y^{}_\nu \right)^{\rm T} C^{}_{BN} \right]  \;,
	\nonumber
	\\
	\mu \frac{{\rm d} C^{}_{HN}}{{\rm d} \mu} &=&  \frac{1}{16\pi^2}  \left[  \left( 2T - \frac{3}{2} g^2_1 - \frac{9}{2}g^2_2  + 12\lambda \right) C^{}_{HN} +  \left( 4C^{}_{HN} - 3\rmi g^{}_1 C^{}_{BN} \right)  Y^\dagger_\nu Y^{}_\nu \right. 
	\nonumber
	\\
	&& + \left. \left( Y^\dagger_\nu Y^{}_\nu \right)^{\rm T} \left( 4C^{}_{HN} + 3\rmi g^{}_1 C^{}_{BN} \right) \right] \;,
\end{eqnarray}
where $T \equiv {\rm Tr} \left( Y^{}_\nu Y^\dagger_\nu + Y^{}_l Y^\dagger_l + 3Y^{}_{\rm u} Y^\dagger_{\rm u} + 3Y^{}_{\rm d}Y^\dagger_{\rm d}  \right)$ and $T^\prime \equiv {\rm Tr} \left[ \left( Y^{}_\nu Y^\dagger_\nu  \right)^2+ \left( Y^{}_l Y^\dagger_l \right)^2+ 3\left( Y^{}_{\rm u}  Y^\dagger_{\rm u}  \right)^2 \right.$ $\left.+ 3 \left( Y^{}_{\rm d}  Y^\dagger_{\rm d}  \right)^2 \right] $ have been introduced. As can be seen from the above results, the three dim-5 operators contribute to the RGEs of some renormalizable couplings. According to the non-renormalization theorem~\cite{Cheung:2015aba,Bern:2019wie,Craig:2019wmo}, the operator $\mathcal{O}^{}_{BN}$ can not be renormalized by the other two since it has the lowest holomorphic and antiholomorphic weights among them. Consequently, the RGE of $C^{}_{BN}$ can not get any contribution from Wilson coefficients $C^{}_5$ and $C^{}_{HN}$, as shown in Eq.~\eqref{eq:rge-wc}. The operators $\mathcal{O}^{}_5$ and $\mathcal{O}^{}_{HN}$ have the same holomorphic and antiholomorphic weights, thus the non-renormalization theorem tells nothing about the RG mixing between them. However, as indicated by Eq.~\eqref{eq:rge-wc}, $\mathcal{O}^{}_5$ does not contribute to the renormalization of $\mathcal{O}^{}_{HN}$ or vice versa. First of all, there is no 1PI diagram renormalizing $\mathcal{O}^{}_{HN}$ (or $\mathcal{O}^{}_5$) with an insertion of $\mathcal{O}^{}_5$ (or $\mathcal{O}^{}_{HN}$) due to field ingredients of operators. More specifically, three fields of the inserted operator have to be in the loop, which is impossible at the one-loop level. One the other hand, one has to check whether $\mathcal{O}^{}_5$ (or $\mathcal{O}^{}_{HN}$) renormalizes $\mathcal{R}^{}_{\ell HN2}$ (or $\mathcal{R}^{}_{\ell HN1}$) since $\mathcal{R}^{}_{\ell HN1}$ and $\mathcal{R}^{}_{\ell HN2}$ can be converted to $\mathcal{O}^{}_5$ and $\mathcal{O}^{}_{HN}$ via EoMs of right-handed neutrinos and lepton doublets, respectively. After calculations, it turns out that $\mathcal{O}^{}_5$ contributes to $\mathcal{R}^{}_{\ell HN1}$ but not to $\mathcal{R}^{}_{\ell HN2}$ (see more details in Appendix~\ref{app:A}), and the case for $\mathcal{O}^{}_{HN}$ is completely opposite. Thus, there is no RG mixing between $\Op^{}_5$ and $\Op^{}_{HN}$.

\section{\label{sec:1}The RGEs for the Seesaw Mechanism}

In the type-I seesaw mechanism with non-degenerate right-handed neutrino masses, one needs to integrate out right-handed neutrinos sequentially at their mass scales (i.e., seesaw scales) when the RG running effects from the GUT scale to the electroweak scale are discussed~\cite{Antusch:2002rr}. Among different seesaw scales, higher mass-dimensional operators, the neutrino Yukawa coupling interactions, and right-handed neutrino(s) $N^{}_{i \rm R}$ (for $i = 1,2, \dots <  n^{}_{\rm max}$ with $n^{}_{\rm max}$ being the initial number of right-handed neutrinos) indeed exist simultaneously, which can be described by the $\nu$SMEFT. At $\mathcal{O} \left( 1/M^{}_i \right) $ with $M^{}_i$ being right-handed neutrino masses, only the unique Weinberg operator $\Op^{}_5 = \overline{\ell^{}_{\rm L}} \widetilde{H} \widetilde{H}^{\rm T} \ell^{\rm c}_{\rm L}$ shows up when right-handed neutrinos are successively integrated out at the tree level and its Wilson coefficient $C^{}_5$ is governed by the matching condition at each seesaw scale~\cite{Chankowski:1993tx,Antusch:2002rr,Broncano:2002rw}. Then, parameters evolve from a superhigh energy scale above seesaw scales (e.g., the GUT scale) to low energy scales with the help of their RGEs. 

For short, EFT($n$) denotes the $\nu$SMEFT with $n$ right-handed neutrinos and in it the Wilson coefficient of the Weinberg operator, the neutrino Yukawa matrix and the right-handed neutrino mass matrix are labeled with a superscript ``$(n)$", i.e., $C^{(n)}_5$, $Y^{(n)}_\nu$ and $M^{(n)}_N$. At the seesaw scale $\mu =M^{}_{n+1}$ where $N^{}_{(n+1)\rm R}$ decouples, the tree-level matching condition for $C^{(n)}_5$ is given by~\footnote{The right-handed neutrinos must be integrated out in their mass eigenstates. However, the running of right-handed neutrino mass matrix does not keep it diagonal though it is initially diagonal. In this case, one needs to transform the right-handed neutrino fields by an unitary matrix $U^{}_N$, namely $N^{}_{\rm R} \to U^\ast_N N^{}_{\rm R}$ to diagonalize the mass matrix, and the neutrino Yukawa coupling matrix is accordingly transformed as $ Y^{}_\nu \to Y^{}_\nu U^\ast_N$. Moreover, at each seesaw scale, $M^{}_i$ deviates its initial value due to running effects. Hereafter, we still use $Y^{}_\nu$ and $M^{}_N$ for those after the diagonalization of $M^{}_N$.}
\begin{eqnarray}
	\left( C^{(n)}_5 \right)^{}_{\alpha\beta} &=& \left( C^{(n+1)}_5 \right)^{}_{\alpha\beta}  +  \left(Y^{(n+1)}_\nu\right)_{\alpha (n+1)} \;M^{-1}_{n+1} \left( Y^{(n+1)}_\nu \right)^{}_{\beta(n+1)} \;\;\;\;
\end{eqnarray}
and $Y^{(n)}_\nu$ is obtained by removing the last column of $Y^{(n+1)}_\nu$ while $ M^{(n)}_N$ is got by removing both the last raw and the last column of $M^{(n+1)}_N$. Therefore, EFT($n^{}_{\rm max}$) is just the full type-I seesaw mechanism with $Y^{(n^{}_{\rm max})}_\nu = Y^{}_\nu$, $M^{(n^{}_{\rm max})}_N = M^{}_N$ and $C^{(n^{}_{\rm max})}_5 = 0$, and EFT($0$) is the SMEFT up to dimension five with $Y^{(0)}_\nu = M^{(0)}_N =0$.

\begin{figure}[t]
	\centering
	\includegraphics[width=0.65\linewidth]{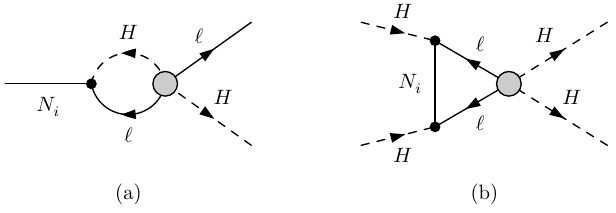}
	\vspace{-0.7cm}
	\caption{1PI diagrams giving new contributions to the relevant RGEs. (a) is for $Y^{}_\nu$ and $C^{}_5$, and (b) for the Higgs quartic coupling $\lambda$.}
	\label{fig:three-point}
\end{figure}

The one-loop RGEs for all couplings in the EFT($0$) and EFT($n^{}_{\rm max}$) have been derived in Refs.~\cite{Chankowski:1993tx,Babu:1993qv,Antusch:2001ck} and~\cite{Casas:1999tp,Antusch:2002rr}, respectively. Those in the EFT($n$) with $n \neq 0, n^{}_{\rm max} $ are more complicated but have been achieved for more than two decades~\cite{Antusch:2002rr}. As we have obtained the complete one-loop RGEs up to dimension five in the $\nu$SMEFT, one can simply get the RGEs for EFT$(n)$ with $n \neq 0, n^{}_{\rm max} $ from those in Eqs.~\eqref{eq:rge-coupling} and ~\eqref{eq:rge-wc} with $\Op^{}_{HN}$ and $\Op^{}_{BN}$ being switched off. Here, we only present those having some discrepancies with the previous results in Ref.~\cite{Antusch:2002rr}, namely
\begin{eqnarray} \label{eq:type-i}
	16\pi^2 \frac{{\rm d} \lambda}{{\rm d} t}  &=& 4\lambda T^{(n)} + \frac{3}{8} (g_1^2+g_2^2)^2 + \frac{3}{4}g^4_2 + 24 \lambda^2  - 3\lambda(g^2_1 + 3g^2_2)  - 2 T^{\prime (n)} + 2{\rm Tr} \left( C^{(n)}_5 Y^{(n)\ast}_\nu  \right. 
	\nonumber
	\\
	&&  \times  \left.  M^{(n)}_N Y^{(n)\dagger}_\nu + Y^{(n)}_\nu M^{(n)\dagger}_N Y^{(n)\rm T}_\nu  C^{(n)\dagger}_5 \right)   \;,
	\nonumber
	\\
	16\pi^2 \frac{{\rm d}  Y^{(n)}_\nu }{{\rm d} t} &=& \frac{1}{2} \alpha^{(n)}_1 Y^{(n)}_\nu  + \frac{3}{2} \left( Y^{(n)}_\nu Y^{(n)\dagger}_\nu - Y^{}_l Y^\dagger_l   \right) Y^{(n)}_\nu   - 3 C^{(n)}_5 Y^{(n)\ast}_\nu M^{(n)}_N  \;,
	\nonumber
	\\
	16\pi^2 \frac{{\rm d} C^{(n)}_5}{{\rm d} t} &=&  \alpha^{(n)}_2 C^{(n)}_5  +  \left(  \frac{7}{2} Y^{(n)}_\nu Y^{(n)\dagger}_\nu - \frac{3}{2}Y^{}_l Y^\dagger_l \right) C^{(n)}_5  + C^{(n)}_5 \left( \frac{7}{2} Y^{(n)}_\nu Y^{(n)\dagger}_\nu  - \frac{3}{2} Y^{}_lY^\dagger_l  \right)^{\rm T} \;
\end{eqnarray}
with $t \equiv \ln (\mu)$ and
\begin{eqnarray}
	\alpha^{(n)}_1  &=& 2 T^{(n)} - \frac{3}{2} g^2_1 - \frac{9}{2} g^2_2  \;,
	\nonumber
	\\
	\alpha^{(n)}_2  &=& 4\lambda + 2T^{(n)} -3g^2_2 \;,
	\nonumber
	\\
	T^{(n)} &=& {\rm Tr} \left( Y^{(n)}_\nu Y^{(n)\dagger}_\nu + Y^{}_l Y^\dagger_l + 3Y^{}_{\rm u} Y^\dagger_{\rm u} + 3Y^{}_{\rm d}Y^\dagger_{\rm d}  \right)\;,
	\nonumber
	\\
	T^{\prime (n)} &=& {\rm Tr} \left[ \left( Y^{(n)}_\nu Y^{(n)\dagger}_\nu  \right)^2+ \left( Y^{}_l Y^\dagger_l \right)^2 + 3\left( Y^{}_{\rm u}  Y^\dagger_{\rm u}  \right)^2 + 3 \left( Y^{}_{\rm d}  Y^\dagger_{\rm d}  \right)^2 \right]  \;.
	\nonumber
\end{eqnarray}
As shown in Eq.~\eqref{eq:type-i}, the Weinberg operator gives contributions to the RGEs of $\lambda$ and $Y^{(n)}_\nu$, which are absent in Ref.~\cite{Antusch:2002rr}, and the factor $7/2$ in front of $Y^{(n)}_\nu Y^{(n)\dagger}_\nu$ instead $1/2$ in the RGE of $C^{(n)}_5$ also contains a new contribution. These discrepancies originate from two one-particle-irreducible (1PI) diagrams shown in Fig.~\ref{fig:three-point}. Diagram (a) is for the renormalization of the $\ell H N$ vertex and gives a contribution to the counterterm of $\Opr^{}_{\ell HN1}$, which is converted to those of $Y^{(n)}_\nu$ and $C^{(n)}_5$ via the EoM of $N^{}_{\rm R}$. Diagram (b) renormalizes Higgs quartic vertex and directly contributes to the counterterm of $\lambda$. However, these two diagrams are not taken into consideration in Ref.~\cite{Antusch:2002rr} since it mainly concentrates on the renormalization of the Weinberg-operator vertex, more specifically, the 1PI diagrams with two lepton and two Higgs external lines. More details about contributions from the two diagrams in Fig.~\ref{fig:three-point} can be found in Appendix~\ref{app:A}. Note that the RGEs in the MSSM~\cite{Antusch:2002rr} do not get such contributions due to the SUSY non-renormalization theorem~\cite{Wess:1973kz,Iliopoulos:1974zv,Grisaru:1979wc}. 

\section{The Effective Neutrino Mass Matrix}\label{sec:eff-mass}

The effective neutrino mass matrix can be defined as $M^{(n)}_\nu = -v^2 \kappa^{(n)}_\nu/2$ with $v$ being the vacuum expectation value of the Higgs field and
\begin{eqnarray}\label{eq:nem}
	\kappa^{(n)}_\nu=  C^{(n)}_5 + \kappa^{(n)}_D 
\end{eqnarray}
where $ \kappa^{(n)}_D  =  Y^{(n)}_\nu M^{(n)-1}_N  Y^{(n)\rm T}_\nu $ results from the mixing among the  light and remaining heavy neutrinos. With the help of RGEs of $ Y^{(n)}_\nu $ and $ M^{(n)}_N$, that of $\kappa^{(n)}_D $ turns out to be
\begin{eqnarray}\label{eq:kappaD}
	16\pi^2 \frac{{\rm d} \kappa^{(n)}_D}{{\rm d} t} &=& \alpha^{(n)}_1  \kappa^{(n)}_D + \beta^{(n)}  \kappa^{(n)}_D  +  \kappa^{(n)}_D \beta^{(n)\rm T}  - 3  Y^{(n)}_\nu Y^{(n)\dagger}_\nu C^{(n)}_5 - 3 C^{(n)}_5  Y^{(n)\ast}_\nu Y^{(n)\rm T}_\nu  \;
\end{eqnarray}
with $ \beta^{(n)} = \left(  Y^{(n)}_\nu Y^{(n)\dagger}_\nu - 3 Y^{}_l Y^\dagger_l  \right) /2$, 
which involves new contributions from the Weinberg operator. However, if one turns to the RGE of $\kappa^{(n)}_\nu$ (or equivalently $M^{(n)}_\nu$), these new contributions in $C^{(n)}_5$ and $ \kappa^{(n)}_D$ are cancelled with each other, leading to the same result for $ \kappa^{(n)}_\nu$ as that in Ref.~\cite{Antusch:2005gp}, namely,
\begin{eqnarray}\label{eq:kappanu}
	16\pi^2 \frac{{\rm d} \kappa^{(n)}_\nu}{{\rm d} t} &=&   \alpha^{(n)}_1  \kappa^{(n)}_\nu  + \beta^{(n)}  \kappa^{(n)}_\nu  + \kappa^{(n)}_\nu \beta^{(n) \rm T}  + \left(\alpha^{(n)}_2 - \alpha^{(n)}_1 \right) C^{(n)}_5 \;.
\end{eqnarray}
Though the counteraction of the new contributions in $ C^{(n)}_5$ and $ \kappa^{(n)}_D$ keeps the RGE of $ \kappa^{(n)}_\nu$ in the same form as that in Ref.~\cite{Antusch:2005gp}, these additional contributions can still affect the running of $ \kappa^{(n)}_\nu$ indirectly via those of $\lambda$, $ C^{(n)}_5$ and $ Y^{(n)}_\nu$, which behave like two-loop effects. To explicitly show their effects on lepton flavor mixing parameters, we adopt the standard parametrization for the lepton mixing matrix $V \equiv U^\dagger_l U^{}_\nu$ with $U^{}_l$ and $U^{}_\nu$ diagonalizing $Y^{}_l$ and $\kappa^{}_\nu$, respectively~\footnote{Due to the existence of $Y^{(n)}_\nu$ in the $Y^{}_l$'s RGE between seesaw scales, $Y^{}_l$ gets non-zero off-diagonal elements during RG running above seesaw scales even it is initially diagonal. On the other hand, the unphysical phase matrix $P^{}_l \equiv {\rm Diag}\{ e^{\rmi \phi^{}_e},  e^{\rmi \phi^{}_\mu}, e^{\rmi \phi^{}_\tau}\}$ is also involved in the RG running but these unphysical phases do not affect the RG running of physical parameters.},
\begin{eqnarray}
	V = \begin{pmatrix} c^{}_{12} c^{}_{13} & s^{}_{12} c^{}_{13} & s^{}_{13} e^{-\rmi\delta} \cr -s^{}_{12} c^{}_{23} - c^{}_{12} s^{}_{13} s^{}_{23} e^{\rmi \delta} & c^{}_{12} c^{}_{23} - s^{}_{12} s^{}_{13}s^{}_{23} e^{\rmi \delta} & c^{}_{13} s^{}_{23} \cr s^{}_{12} s^{}_{23} - c^{}_{12} s^{}_{13} c^{}_{23} e^{\rmi \delta} & -c^{}_{12} s^{}_{23} - s^{}_{12} s^{}_{13} c^{}_{23} e^{\rmi \delta} & c^{}_{13} c^{}_{23}  \end{pmatrix} \begin{pmatrix} e^{\rmi \rho} & & \cr & e^{\rmi \sigma} & \cr & & 1 \end{pmatrix}
\end{eqnarray}
with $s^{}_{ij} \equiv \sin\theta^{}_{ij}$ and $c^{}_{ij} \equiv \cos\theta^{}_{ij}$ for $ij=12,13,23$.
\begin{figure}[t]
	\centering
	\includegraphics[width=0.7\linewidth]{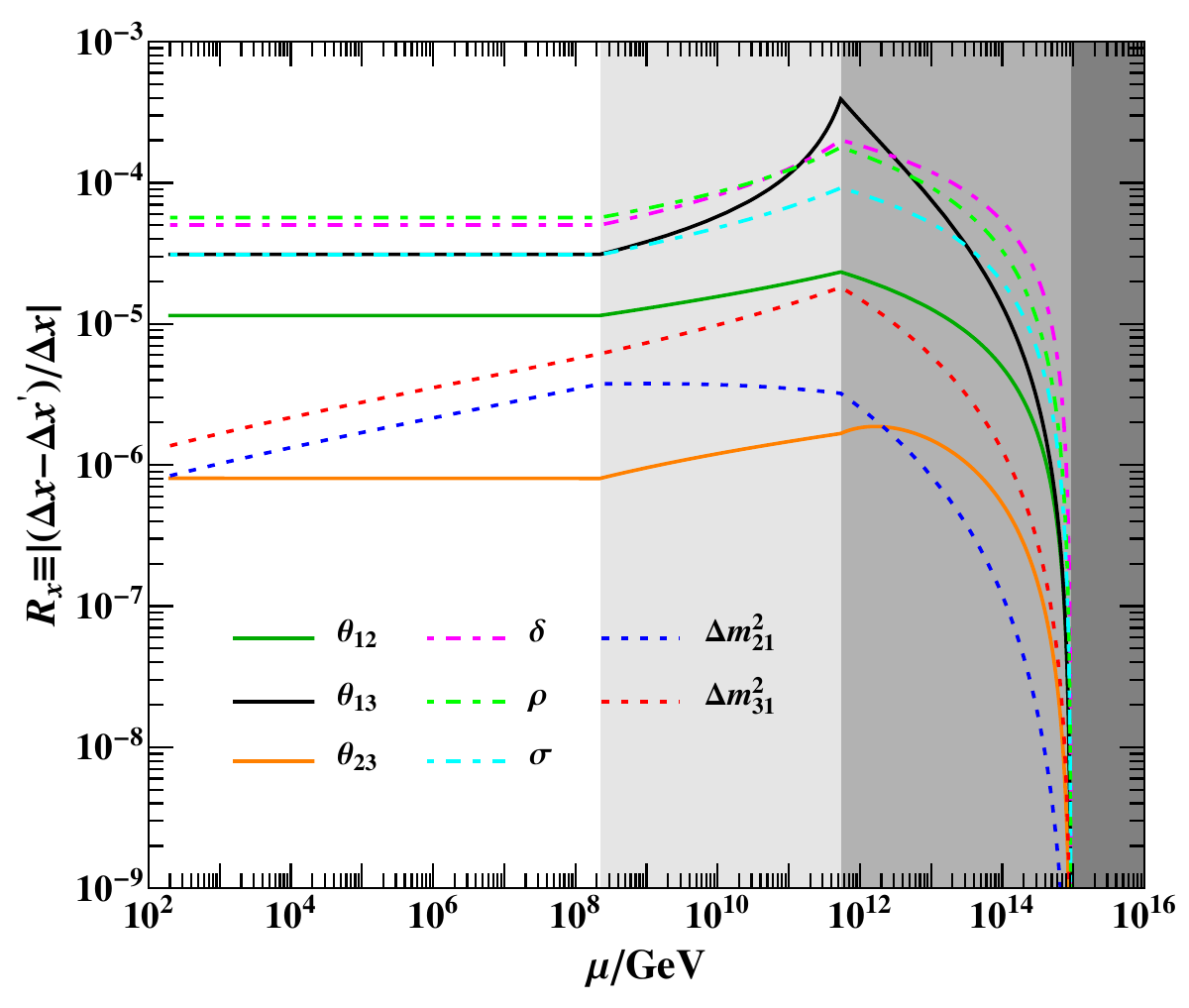}
	\vspace{-0.5cm}
	\caption{The relative size of corrections to neutrino oscillation parameters from RGEs involving the new contributions, i.e., $\Delta x^\prime$ with respect to those without them, i.e., $\Delta x$, with $x^{(\prime)}$ being the neutrino oscillation parameters. Each gray range stands for a valid energy interval of the corresponding EFT.}
	\label{fig:deviation}
\end{figure}
Fig.~\ref{fig:deviation} shows the relative size of RG corrections to the two squared neutrino mass differences, three mixing angles and three CP-violating phases from additional terms with respect to those without them, where the values of right-handed neutrino masses are  $\left(M^{}_1, M^{}_2, M^{}_3\right) = \left( 2.215 \times 10^8, 5.382 \times 10^{11},  9.620 \times 10^{14} \right) $ GeV at the initial scale $\mu = \Lambda^{}_{\rm GUT} \sim 1\times 10^{16}$ GeV
and those for other couplings are chosen to reproduce the values of physical parameters (such as lepton and quark masses and mixing parameters) at the electroweak scale. Note that each right-handed neutrino decouples at a scale slightly lower than its initial mass, namely $\left( 2.215 \times 10^8, 5.381 \times 10^{11}, 9.332\times 10^{14} \right)$ GeV for $\left( N^{}_1, N^{}_2, N^{}_3 \right)$ due to RG running of their masses~\cite{Antusch:2002rr,Antusch:2005gp}. As seen from Fig.~\ref{fig:deviation}, relative discrepancies appear below the highest seesaw scale $\mu = M^{}_3$, and they are pretty small and around $\mathcal{O} (10^{-6}) $ - $ \mathcal{O} (10^{-4})$. As discussed below Eq.~\eqref{eq:kappanu}, such small discrepancies are indirectly caused by the one-loop RG running of $\lambda$, $C^{(n)}_5$ and $Y^{(n)}_\nu$ and equivalent to two-loop effects. As a result, the smallness of those discrepancies between flavor mixing parameters is robust against initial inputs. However, those additional contributions do influence the one-loop RG running behaviors of $\lambda$ and $ Y^{(n)}_\nu$, especially the flavor structure of the latter one, and their effects could be significant but depend on initial inputs. In general, those additional contributions should not be ignored if one concerns about parameter space of the full theory or phenomenologies directly related to $Y^{}_\nu$, such as leptogenesis~\cite{Fukugita:1986hr}.

\section{The Determinant of the Effective Neutrino Mass Matrix}

Taking advantage of Eq.~\eqref{eq:kappanu}, one can achieve
\begin{eqnarray}\label{eq:det}
	16\pi^2 \frac{{\rm d Det}(  \kappa^{(n)}_\nu )}{{\rm d} t} &=&  \left(3 \alpha^{(n)}_1 + 2 {\rm Tr}\beta^{(n)} \right) {\rm Det}( \kappa^{(n)}_\nu )  + \left( \alpha^{(n)}_2 -  \alpha^{(n)}_1 \right){\rm Tr} \left( {\rm Adj} ( \kappa^{(n)}_\nu) C^{(n)}_5 \right)  \;,
\end{eqnarray}
where ``Det'' and ``Adj" denote the determinant and the adjugate matrix, and the following formulae are exploited, namely
\begin{eqnarray}
	\frac{{\rm d} {\rm Det} (M)}{{\rm d} x} &=& {\rm Tr} \left( {\rm Adj} (M) \frac{{\rm d}M}{{\rm d} x} \right) \;,\quad \nonumber
	\\
	M {\rm Adj} (M) &=& {\rm Adj} (M) M =  {\rm Det} (M) \mathbf{1}
\end{eqnarray}
with $\mathbf{1}$ being an identity matrix of the same dimension as the square matrix $M$. 

\subsection{Above the Highest or below the Lowest Seesaw Scale}
In the case above the highest or below the lowest seesaw scale ( i.e., $n = n^{}_{\rm max}$ or $0$), starting with Eq.~\eqref{eq:det}, the determinant of $\kappa^{(n)}_\nu$ is easily found to be
\begin{eqnarray}\label{eq:det-int}
	{\rm Det} [ \kappa^{(n)}_\nu (t) ] =  I^{(n)}_{\rm Int} \left( t\right) \cdot {\rm Det} [ \kappa^{(n)}_\nu (t^{}_0) ] \;,
\end{eqnarray}
in which $t^{}_{0}=\ln (\mu^{}_{0})$ is the initial energy scale in each energy interval and 
\begin{eqnarray}
	I^{(0)}_{\rm Int} \left(t\right) &=& \exp \left[ \frac{1}{16\pi^2} \int^{t}_{t^{}_0} \left(3 \alpha^{(0)}_2 + 2 {\rm Tr} \beta^{(0)} \right) {\rm d} t^\prime   \right] \;,
	\nonumber
	\\
	I^{(n^{}_{\rm max})}_{\rm Int} \left(t\right) &=& \exp \left[  \frac{1}{16\pi^2} \int^{t}_{t^{}_0} \left(3 \alpha^{(n^{}_{\rm max})}_1 + 2 {\rm Tr}\beta^{(n^{}_{\rm max})}\right)  {\rm d} t^\prime  \right] \;.\;\;\;\;\;
\end{eqnarray}
The above results are consistent with those in Refs.~\cite{Mei:2003gn,Zhu:2018dvj,Zhang:2020lsd}. Note that for a complex symmetric matrix, its determinant is also complex in general.
Let $\chi^{(n)}_i$ with $i=1,2,3$ be the singular values of $ \kappa^{(n)}_\nu$ and $\varphi^{(n)}$ be the argument of ${\rm Det} ( \kappa^{(n)}_\nu)$, then Eq.~\eqref{eq:det-int} leads us to
\begin{eqnarray}\label{eq:singular-ab}
	\chi^{(n)}_1 \hspace{-2pt}\left(t\right)  \chi^{(n)}_2 \hspace{-2pt}\left(t\right)  \chi^{(n)}_3 \hspace{-2pt}\left(t\right)  &=& I^{(n)}_{\rm Int} \hspace{-2pt}\left( t\right) \cdot	\chi^{(n)}_1 \hspace{-2pt}\left(t^{}_0\right)  \chi^{(n)}_2 \hspace{-2pt}\left(t^{}_0\right) \chi^{(n)}_3 \hspace{-2pt}\left(t^{}_0\right) \;,\quad
	\nonumber
	\\
	\varphi^{(n)}\left( t \right) &=& \varphi^{(n)}\left( t^{}_0 \right) \;.
\end{eqnarray}
One the other hand, directly starting with Eq.~\eqref{eq:kappanu}, one can derive 
\begin{eqnarray}\label{eq:singular-ab-1}
	 &&\chi^{(n)}_i \left( t \right) = I^{(n)}_{i,\rm Int} \left( t\right) \cdot \chi^{(n)}_i \left( t^{}_0 \right)  \;,
	 \nonumber
	 \\
	 &&  {\rm Det} [ U^{(n)}_\nu (t) ] =  {\rm Det} [ U^{(n)}_\nu (t^{}_0) ] \;,
\end{eqnarray}
for $n = 0$ and $n^{}_{\rm max}$, where
\begin{eqnarray}
	I^{(0)}_{i,\rm Int} \left( t\right) &=& \exp \left[  \frac{1}{16\pi^2} \int^{t}_{t^{}_0} \left( \alpha^{(0)}_2+ 2  \beta^{\prime (0)}_{ii}\; \right) {\rm d} t^\prime   \right] \;,
	\nonumber
	\\
	I^{(n^{}_{\rm max})}_{i, \rm Int} \left(t\right) &=& \exp \left[  \frac{1}{16\pi^2} \int^{t}_{t^{}_0} \left( \alpha^{(n^{}_{\rm max})}_1 + 2  \beta^{\prime (n^{}_{\rm max})}_{ii}  \; \right)  {\rm d} t^\prime   \right] \; \;\;\;\;\;\;
\end{eqnarray}
and $\beta^{\prime (n)} = U^{(n)\dagger}_\nu \beta^{(n)} U^{(n)}_\nu$ with $U^{(n)}_\nu $ being the unitary matrix diagonalizing $\kappa^{(n)}_\nu $, i.e., $U^{(n)\dagger}_\nu \kappa^{(n)}_\nu U^{(n)\ast}_\nu = {\rm Diag} \{ \chi^{(n)}_1,  \chi^{(n)}_2,  \chi^{(n)}_3 \}$. Some comments on the above results are given in order.
\begin{itemize}
	\item  It is easy to check that $I^{(n)}_{\rm Int} \left( t\right) = \prod^{3}_{i=1} I^{(n)}_{i,\rm Int} \left( t\right) $ holds and hence the first equation in Eq.~\eqref{eq:singular-ab} can be easily produced by that in Eq.~\eqref{eq:singular-ab-1}. 
	
	\item The argument $ \varphi^{(n)}$ of ${\rm Det} ( \kappa^{(n)}_\nu )$ is equivalent to that of $[ {\rm Det} ( U^{(n)}_\nu )]^2$, thus the second equation in Eq.~\eqref{eq:singular-ab-1} easily leads to that in Eq.~\eqref{eq:singular-ab}. As indicated, $ \varphi^{(n)}$ is  invariant during RG running above the highest or below the lowest seesaw scale. The specific form of $ \varphi^{(n)}$ depends on the parametrization of $ U^{(n)}_\nu $ and usually involves unphysical physics~\cite{Xing:2005fw,Ohlsson:2012pg,Zhang:2020lsd}. 
	
	\item As seen from Eqs.~\eqref{eq:singular-ab} and \eqref{eq:singular-ab-1}, both ${\rm Det} ( \kappa^{(n)}_\nu )$ and the singular values $\chi^{(n)}_i$ (for $n = 0, n^{}_{\rm max}$) are proportional to their initial values.  Consequently, a non-zero determinant or singular value can be never induced by running of the zero one, or vice versa, which has been found and discussed in Refs.~\cite{Antusch:2005gp,Davidson:2006tg,Xing:2020ezi,Zhang:2020lsd}. Moreover, the matrix rank of $ \kappa^{(n)}_\nu$ remains unchanged during running above the highest or below the lowest seesaw scale.
\end{itemize}

\subsection{Between Seesaw Scales}

However, cases among seesaw scales become much more complicated and are unclear so far. Very recently, Ref.~\cite{Benoit:2022ohv} made an attempt and concluded that the neutrino mass matrix remains rank degenerate among seesaw scales in the minimal type-I seesaw mechanism but the answer is undefined and depends on the kernel solutions of RGEs in the canonical one with the neutrino mass matrix being rank degenerate above the highest seesaw scale. These two conclusions are achieved based on the previous results for the RGEs~\cite{Antusch:2002rr}, where some additional contributions from the Weinberg operator among seesaw scales are overlooked. Taking into account the updated RGE for $ Y^{(n)}_\nu$ in Eq.~\eqref{eq:type-i}, the kernel function for $Y^{(n)}_\nu$ playing an essential role in the derivation is no longer valid and therefore neither are the above conclusions. To our best knowledge, there is no strict proof or solid conclusion on this issue so far even it is widely expected that the lightest neutrino's mass can not be generated via the one-loop RG running effects if it is initially zero in the seesaw mechanism. In this work, we attempt to offer a strict proof and make a definite conclusion on this issue.

Compared with the case above the highest or below the lowest seesaw scale, the difficulty appearing in cases  among seesaw scales is the entanglement between $\kappa^{(n)}_\nu = C^{(n)}_5 + \kappa^{(n)}_D$ and $C^{(n)}_5$, as shown in Eq.~\eqref{eq:kappanu} or~\eqref{eq:det}. If we want to determine the determinant of $ \kappa^{(n)}_\nu$ with the help of Eq.~\eqref{eq:det}, we have to figure out what ${\rm Tr} [  {\rm Adj} ( \kappa^{(n)}_\nu) C^{(n)}_5 ]$ is and how it is related to ${\rm Det} ( \kappa^{(n)}_\nu)$. Fortunately, for three generations of active neutrinos (i.e., $\kappa^{(n)}_\nu$, $\kappa^{(n)}_D$ and $C^{(n)}_5$ are all $3\times3$ matrices), we can obtain 
\begin{eqnarray}\label{eq:dett1}
	{\rm Tr} \left[ {\rm Adj} ( \kappa^{(n)}_\nu) C^{(n)}_5 \right] &=& {\rm Det} ( \kappa^{(n)}_\nu) + 2 {\rm Det} ( C^{(n)}_5) - {\rm Det} ( \kappa^{(n)}_D )  + {\rm Tr} \left[  {\rm Adj} ( C^{(n)}_5 ) \kappa^{(n)}_D \right] \;,
	\\
	\label{eq:dett2}
	{\rm Tr} \left[ {\rm Adj} ( \kappa^{(n)}_\nu ) C^{(n)}_5 \right] &=& 2 {\rm Det} ( \kappa^{(n)}_\nu ) +  {\rm Det} ( C^{(n)}_5 ) - 2{\rm Det} (\kappa^{(n)}_D)  - {\rm Tr} \left[ {\rm Adj} ( \kappa^{(n)}_D ) C^{(n)}_5 \right] \;,
\end{eqnarray}
which originate from linear algebra and are independent of the energy scale $\mu$. The derivation for the above two relations can be found in Appendix~\ref{app:adj}.
 
As indicated by Eq.~\eqref{eq:type-i}, the RGE of $ C^{(n)}_5 $ has the same form as that of $ \kappa^{(i)}_\nu $ with $i=0,n^{}_{\rm max}$. Therefore, the conclusions on the determinant and the matrix rank of $ \kappa^{(i)}_\nu $ are also valid for  $C^{(n)}_5$. Namely,  ${\rm Det} ( C^{(n)}_5)$ is proportional to its initial value at the seesaw scale $\mu= M^{}_{n+1}$ and the rank of $ C^{(n)}_5$, i.e., ${\rm Rank} (C^{(n)}_5)$ is invariant and only depends on the initial one at $\mu= M^{}_{n+1}$. It is not easy to determine ${\rm Det} ( \kappa^{(n)}_D) $ from Eq.~\eqref{eq:kappaD} but $\kappa^{(n)}_D$ is defined as a combination of $Y^{(n)}_\nu$ and $M^{(n)}_N$ at any energy scale and hence its rank is governed by the minimal one between those of $ Y^{(n)}_\nu$ and $ M^{(n)}_N$, i.e., ${\rm Rank} ( \kappa^{(n)}_D) = {\rm min} \{ {\rm Rank} (Y^{(n)}_\nu), {\rm Rank} ( M^{(n)}_N )  \} $. Moreover, ${\rm Adj}(M) = \mathbf{0}$ holds if $M$ is a $k\times k$ matrix and satisfies ${\rm Rank}(M) \leq k - 2$. Now that we have all we can have in general, we go to the minimal type-I seesaw mechanism and the canonical one, respectively. In both cases, we consider the matrix rank of the effective neutrino mass matrix is initially not small than two. In other worlds, at most one neutrino is initially massless.

\subsubsection{The Minimal Type-I Seesaw Mechanism}

There are only two right-handed neutrinos (i.e., $n^{}_{\rm max} = 2$) in the minimal type-I seesaw mechanism, therefore, ${\rm Rank} [ Y^{(2)}_\nu (\mu) ] = 2$ and ${\rm Det} [ \kappa^{(2)}_\nu(\mu) ] = 0 $ hold for $\mu \geq M^{}_2$. At $\mu = M^{}_2$, the tree-level matching condition gives us $(C^{(1)}_5)^{}_{\alpha\beta}  = ( Y^{(2)}_\nu )^{}_{\alpha 2} ( Y^{(2)}_\nu )^{}_{\beta 2}/M^{}_2 $, leading to ${\rm Rank} [ C^{(1)}_5 (\mu) ] =  {\rm Rank} [ C^{(1)}_5 (M^{}_2) ] = 1$ with $M^{}_1 \leq \mu \leq M^{}_2$. $Y^{(1)}_\nu (\mu)$ is a column vector and satisfies ${\rm Rank} [ Y^{(1)}_\nu (\mu) ] = 1$. This indicates that $\kappa^{(1)}_D (\mu) =  Y^{(1)}_\nu (\mu) Y^{(1)\rm T}_\nu (\mu)/M^{}_1 (\mu)$ is also a rank one matrix, namely ${\rm Rank} [ \kappa^{(1)}_D (\mu) ] = 1$. Therefore, in $M^{}_1 \leq \mu \leq M^{}_2$, we have ${\rm Rank} [ C^{(1)}_5   \left( \mu \right) ] = {\rm Rank} [ \kappa^{(1)}_D \left( \mu \right)] = 1$, which results in ${\rm Det} [ \kappa^{(1)}_\nu \left( \mu \right) ] = 0$ with the help of Eq.~\eqref{eq:dett1} or the fact that ${\rm Rank} [ \kappa^{(1)}_\nu \left( \mu \right) ] \leq {\rm Rank} [C^{(1)}_5 \left( \mu \right) ] + {\rm Rank} [\kappa^{(1)}_D \left( \mu \right) ] = 2$. As a result, the determinant of $ \kappa^{(1)}_\nu \left( \mu \right) $ remains zero in  $M^{}_1 \leq \mu \leq M^{}_2$. One can check that this behavior is consistent with Eq.~\eqref{eq:det}.

It comes to the conclusion that in the minimal type-I seesaw mechanism, the determinant of $\kappa^{}_\nu$ is always zero in the whole energy range. This means that the lightest neutrino remains massless even the RG running with threshold effects from the GUT scale to the electroweak scale are taking into account.

\subsubsection{The Canonical Seesaw Mechanism}

The case in the canonical seesaw mechanism (i.e., $n^{}_{\rm max} = 3$) becomes more complicated. We discuss it along the decreasing energy scale:
\begin{itemize}
	\item $\mu \geq M^{}_3$, ${\rm Rank} [ \kappa^{(3)}_\nu (\mu) ] = {\rm Rank} [ Y^{(3)}_\nu (\mu)] \leq 3$ hold;
	
	\item At $\mu = M^{}_3$, $ (C^{(2)}_5 )^{}_{\alpha\beta}  = ( Y^{(3)}_\nu )^{}_{\alpha 3} ( Y^{(3)}_\nu )^{}_{\beta 3} /M^{}_3 $ causes ${\rm Rank} [ C^{(2)}_5 (M^{}_3 )] = 1$, and we have $ {\rm Det} [ \kappa^{(2)}_\nu (M^{}_3) ]  = {\rm Det} [ \kappa^{(3)}_\nu (M^{}_3) ] $ due to $\kappa^{(2)}_\nu (M^{}_3) = \kappa^{(3)}_\nu (M^{}_3)$;
	
	\item In $M^{}_2 \leq \mu \leq M^{}_3$, the invariance of $C^{(2)}_5 (\mu)$'s rank gives ${\rm Rank} [ C^{(2)}_5 (\mu) ] = {\rm Rank} [ C^{(2)}_5 (M^{}_3) ] = 1$, and ${\rm Rank} [ \kappa^{(2)}_D (\mu) ] =  {\rm Rank} [ Y^{(2)}_\nu (\mu) ] \leq 2$ holds since $ Y^{(2)}_\nu (\mu)$ is a $3\times2$ matrix. Consequently, neither $C^{(2)}_5 (\mu) $ nor $\kappa^{(2)}_D (\mu)$ has a full rank resulting in ${\rm Det} [ C^{(2)}_5 (\mu) ] = {\rm Det} [ \kappa^{(2)}_D(\mu) ] = 0$. Additionally, ${\rm Rank} [ C^{(2)}_5 (\mu) ] = 1$ leads to ${\rm Adj} [ C^{(2)}_5 (\mu) ] = 0$. From Eq.~\eqref{eq:dett1}, we obtain ${\rm Tr} \left(  {\rm Adj} [ \kappa^{(2)}_\nu \left( \mu \right) ] C^{(2)}_5 \left( \mu \right) \right) = {\rm Det} [ \kappa^{(2)}_\nu \left( \mu \right) ]$. Substituting it into Eq.~\eqref{eq:det}, one has ${\rm d Det} (\kappa^{(2)}_\nu)/{\rm d} t \propto {\rm Det} ( \kappa^{(2)}_\nu )$ and this indicates $ {\rm Det} [ \kappa^{(2)}_\nu (M^{}_2) ]  \propto  {\rm Det} [ \kappa^{(2)}_\nu (M^{}_3) ] $;
	
	\item At $\mu = M^{}_2$, with the tree-level matching condition $ (C^{(1)}_5 )^{}_{\alpha\beta}  = (C^{(2)}_5 )^{}_{\alpha\beta} +( Y^{(2)}_\nu )^{}_{\alpha 2} ( Y^{(2)}_\nu )^{}_{\beta 2}/M^{}_2 $, one gets ${\rm Rank} [ C^{(1)}_5 (M^{}_2) ] \leq  {\rm Rank} [ C^{(2)}_5 (M^{}_2) ] + {\rm Rank} [ {\bf y}^{(2)}_2 (M^{}_2) {\bf y}^{(2)\rm T}_2 (M^{}_2)/M^{}_2] = 2$ with $( {\bf y}^{(2)}_{2} )^{}_\alpha = (Y^{(2)}_\nu   )^{}_{\alpha 2}$ being a $3\times1$ column vector. Moreover, $ {\rm Det} [ \kappa^{(1)}_\nu (M^{}_2) ]  = {\rm Det} [ \kappa^{(2)}_\nu (M^{}_2) ] $ holds;
	
	\item In $M^{}_1 \leq \mu \leq M^{}_2$, ${\rm Rank} [ C^{(1)}_5 (\mu) ] = {\rm Rank} [ C^{(1)}_5 (M^{}_2) ] \leq 2$ induces ${\rm Det} [ C^{(1)}_5 (\mu) ] = 0$, and ${\rm Rank} [ \kappa^{(1)}_D (\mu) ] =  {\rm Rank} [ Y^{(1)}_\nu (\mu) ] = 1 $ gives ${\rm Det} [\kappa^{(1)}_D (\mu) ] = {\rm Adj} [ \kappa^{(1)}_D (\mu) ] = 0$. With the help of Eq.~\eqref{eq:dett2}, one obtains ${\rm Tr} \left(  {\rm Adj} [ \kappa^{(1)}_\nu \left( \mu \right) ] C^{(1)}_5 \left( \mu \right) \right) = 2{\rm Det} [ \kappa^{(1)}_\nu \left( \mu \right) ]$ and subsequently
    $ {\rm Det} [ \kappa^{(1)}_\nu (M^{}_1) ]  \propto  {\rm Det} [ \kappa^{(1)}_\nu (M^{}_2) ]$;
    
    \item At $\mu = M^{}_1$, we have $ ( \kappa^{(0)}_\nu )^{}_{\alpha\beta} = (C^{(0)}_5 )^{}_{\alpha\beta}  = (C^{(1)}_5 )^{}_{\alpha\beta} +( Y^{(1)}_\nu )^{}_{\alpha 1} ( Y^{(1)}_\nu )^{}_{\beta 1}/M^{}_1 $, $Y^{(0)}_\nu = M^{(0)}_N = 0$, and $ {\rm Det} [ \kappa^{(0)}_\nu (M^{}_1) ]  = {\rm Det} [ \kappa^{(1)}_\nu (M^{}_1) ] $;
    
    \item $\mu \leq M^{}_1$, we have ${\rm Tr} \left(  {\rm Adj} [ \kappa^{(0)}_\nu \left( \mu \right) ] C^{(0)}_5 \left( \mu \right) \right) = 3 {\rm Det} [ \kappa^{(0)}_\nu \hspace{-0.05cm}\left( \mu \right) ]$ and 
    $ {\rm Det} [ \kappa^{(0)}_\nu (\Lambda^{}_{\rm EW}) ]  \propto  {\rm Det} [ \kappa^{(0)}_\nu (M^{}_1) ]$;
\end{itemize}
From the above discussion and the formula in Eq.~\eqref{eq:det}, one gets ${\rm Tr} [  {\rm Adj} (\kappa^{(n)}_\nu ) C^{(n)}_5 ] = (3-n) {\rm Det} ( \kappa^{(n)}_\nu )$ and 
\begin{eqnarray} \label{eq:det3}
	16\pi^2\frac{{\rm d Det}( \kappa^{(n)}_\nu )}{{\rm d}t} &=& \left[ 3 \alpha^{(n)}_1 + (3 - n)\left(  \alpha^{(n)}_2 - \alpha^{(n)}_1 \right) + 2 {\rm Tr} \beta^{(n)}  \right] {\rm Det}( \kappa^{(n)}_\nu )
\end{eqnarray}
with $n=0,1,2,3$. Integrating both sides of Eq.~\eqref{eq:det3} in each energy interval $ t^{}_n \leq t \leq t^{}_{n+1}$, one achieves
\begin{eqnarray}\label{eq:detint-p}
	{\rm Det} [ \kappa^{(n)}_\nu (t) ] = I^{\prime (n)}_{\rm Int} \left( t, t^{}_{n+1} \right) \cdot {\rm Det} [ \kappa^{(n)}_\nu (t^{}_{n+1}) ]
\end{eqnarray}
where $t^{}_{i} \equiv \ln (\mu^{}_{i})$ is introduced with $\mu^{}_i = M^{}_i$ (for $i=1,2,3$) and $\mu^{}_4 = \Lambda^{}_{\rm GUT}$, and
\begin{eqnarray}\label{eq:int-f1}
	I^{\prime (n)}_{\rm Int} \left(t,t^{}_{n+1}\right) &=& \exp \left\{ \frac{1}{16\pi^2} \int^{t}_{t^{}_{n+1}} \left[ n \alpha^{(n)}_1 + (3 - n) \alpha^{(n)}_2 + 2 {\rm Tr} \beta^{(n)} \right]  {\rm d} t^\prime  \vphantom{\int^{t}_{t^{}_{n+1}} } \right\} \;.
\end{eqnarray}
For the dependence of ${\rm Det} ( \kappa^{}_\nu) $ on its initial value at $\mu_4 = \Lambda^{}_{\rm GUT}$, one can obtain a piecewise function, that is 
\begin{eqnarray}\label{eq:detint}
	{\rm Det} ({\kappa^{}_\nu} (t) ) = {I^{}_{\rm Int}} \left( t \right) \cdot {\rm Det} (\kappa^{}_\nu (t^{}_4) )
\end{eqnarray}
with
\begin{eqnarray}\label{eq:int-f2}
	{I^{}_{\rm Int}} \left(t \right) = \left\{ \begin{array}{lr} {I^{\prime (3)}_{\rm Int}} \left( t,t^{}_4 \right) \;, & t^{}_3< t \leq t^{}_4 \\[0.15cm]  {I^{\prime (2)}_{\rm Int}} \left( t,t^{}_3 \right)  {I^{\prime (3)}_{\rm Int}} \left( t^{}_3,t^{}_4 \right) \;,  & t^{}_2< t \leq t^{}_3 \\[0.15cm] {I^{\prime (1)}_{\rm Int}} \left( t,t^{}_2 \right)  \displaystyle\prod_{i=2}^{3}{I^{\prime (i)}_{\rm Int}} \left( t^{}_i,t^{}_{i+1} \right) \;,  & t^{}_1< t \leq t^{}_2 \\[0.15cm] I^{\prime (0)}_{\rm Int} \left( t,t^{}_1 \right)  \displaystyle\prod_{i=1}^{3}{I^{\prime (i)}_{\rm Int}} \left( t^{}_i,t^{}_{i+1} \right)  \;, & t \leq t^{}_1 \end{array} \right. ,\;\;\;
\end{eqnarray}
where the continuity of the determinant at each seesaw scale has been exploited. As can be seen in Eq.~\eqref{eq:detint}, the determinant  ${\rm Det} ({\kappa^{}_\nu})$ is always proportional to its initial value at $\mu = \Lambda^{}_{\rm GUT}$ even threshold effects among different seesaw scales are included. Therefore, if the determinant is initially zero, it remains vanishing against the RG running from the GUT scale to the electroweak scale. This indicates that the initially massless neutrino can not get a non-zero mass via one-loop RG-running effects. But it can when the two-loop effects are introduced. For instance, there is a two-loop contribution in form of $\left( Y^{}_l Y^\dagger_l \right) C^{}_5 \left( Y^{}_l Y^\dagger_l \right)^{\rm T}$ in the RGE of $C^{}_5$~\cite{Davidson:2006tg,Xing:2020ezi,Ibarra:2024in}, which leads to a nontrivial term, i.e, ${\rm Tr} \left[ {\rm Adj} ( \kappa^{}_\nu ) \left( Y^{}_l Y^\dagger_l \right) \kappa^{}_\nu \left( Y^{}_l Y^\dagger_l \right)^{\rm T} \right] $ in the RGE of ${\rm Det} \left(\kappa^{}_\nu\right)$ below seesaw scales. This term is generally not proportional to ${\rm Det} ({\kappa^{}_\nu})$ due to nontrivial insertions of $Y^{}_l Y^\dagger_l$ and hence makes a non-zero contribution to ${\rm Det} ({\kappa^{}_\nu})$ even if the latter is initially vanishing. It would be much more transparent if one looks at this two-loop effect on the initially vanishing singular value $\chi^{}_1$ or $\chi^{}_3$ of $\kappa^{}_\nu$. As shown in Refs.~\cite{Davidson:2006tg,Xing:2020ezi,Ibarra:2024in}, $\chi^{}_1$ or $\chi^{}_3$ gets non-vanishing radiative corrections approximately proportional to the other two non-zero singular values, namely, $\chi^{}_1 \propto \sum^{}_{i=2,3} \chi^{}_i {\rm Re} \left( V^\ast_{\tau 1} V^{}_{\tau i} \right)^2 $ or $\chi^{}_3 \propto \sum^{}_{i=1,2} \chi^{}_i {\rm Re} \left( V^\ast_{\tau 3} V^{}_{\tau i} \right)^2$. These discussions are also applicable to the two-loop RG effects on the right-handed neutrino mass matrix $M^{}_N$ above seesaw scales, where the two-loop RGE of $M^{}_N$ has a similar rank-increase term $\left( Y^\dagger_\nu Y^{}_\nu \right)^{\rm T} M^{}_N \left( Y^\dagger_\nu Y^{}_\nu \right) $ (see Refs.~\cite{Ibarra:2020eia,Bonilla:2020sne} for more details). Additionally, some discussions about two-loop effects induced by the exchange of two W bosons on light neutrino masses in the framework with extra generation(s) and right-handed neutrino(s) can be found in Refs.~\cite{Babu:1988ig,Grimus:1989pu,Aparici:2011nu,Schmidt:2011jp,Aparici:2012vx}.

\begin{figure}
	\centering
	\subfigure[]{
		\includegraphics[width=0.48\linewidth]{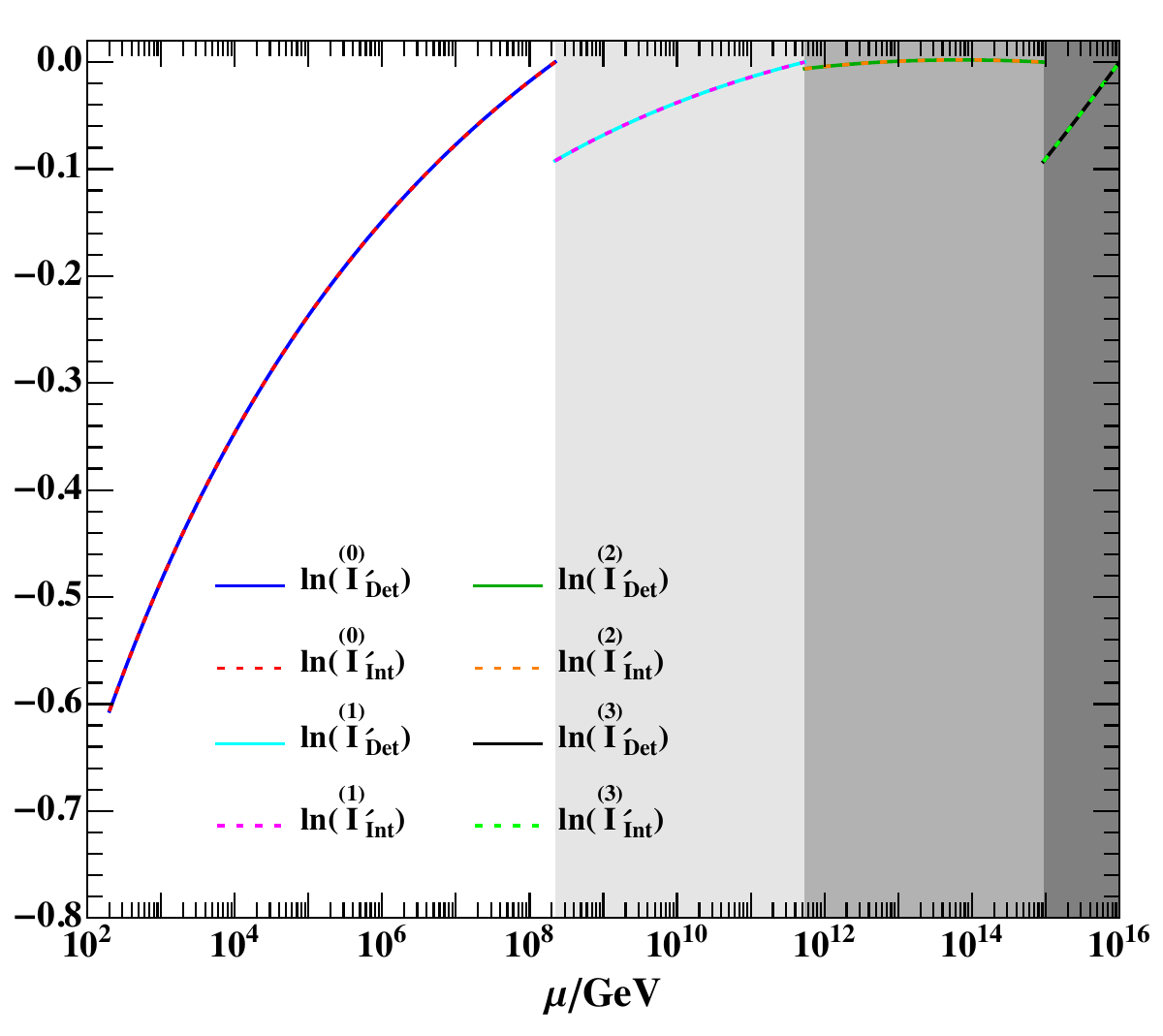}
	}
	\subfigure[]{
		\includegraphics[width=0.48\linewidth]{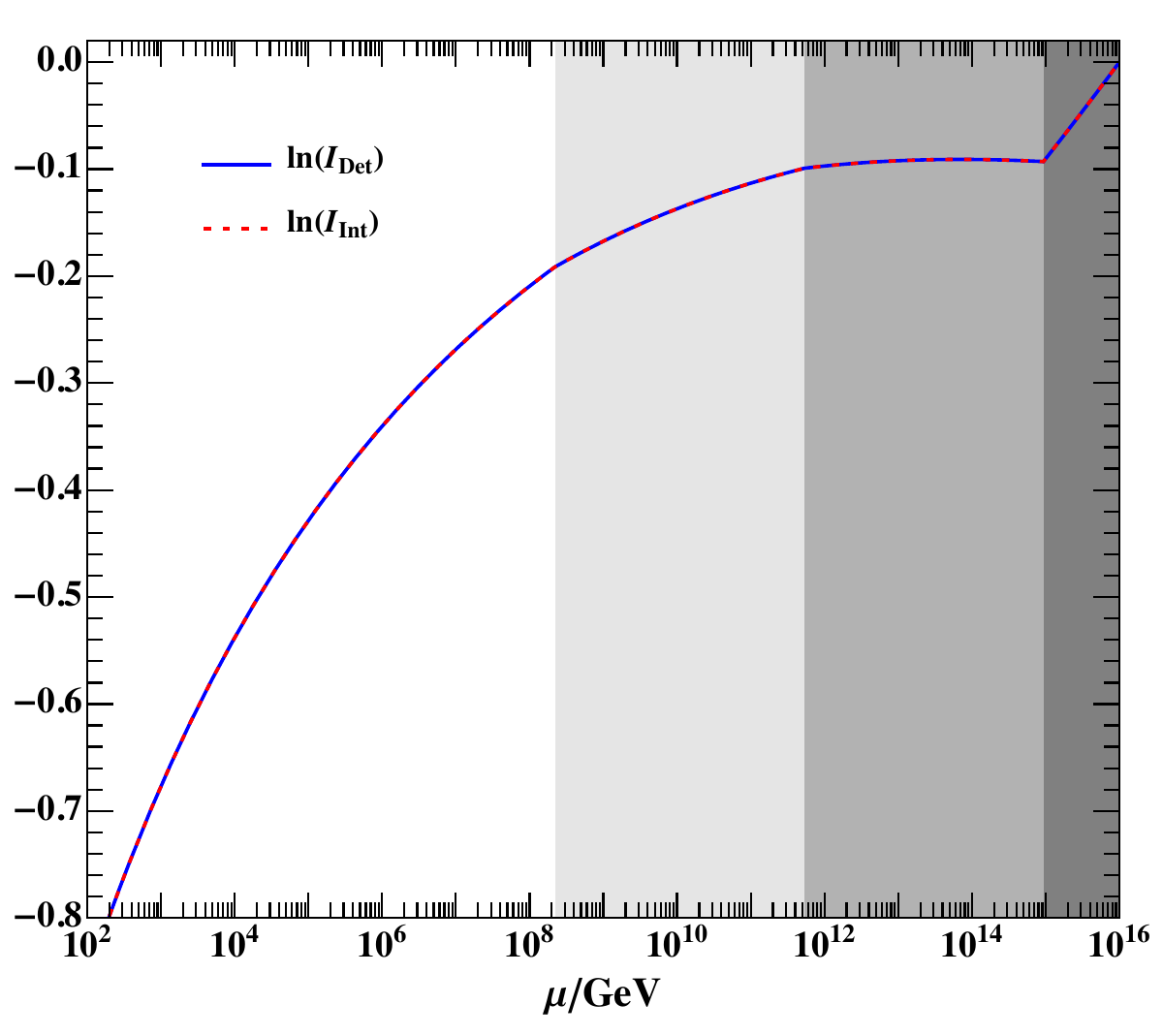}
	}
	\vspace{-0.3cm}
	\caption{The RG running of the  non-vanishing determinant  against the energy scale. (a) shows the ratio of the running determinant to its initial value in each energy interval, and (b) illustrates that to the initial determinant at the GUT scale. $I^{\prime (n)}_{\rm Det} (t)$  and $ I^{}_{\rm Det} (t) $ are directly calculated from $\kappa^{(n)}_\nu$ and $\kappa^{}_\nu$ via $ I^{\prime (n)}_{\rm Det} (t)  = 	{\rm Det} ( \kappa^{(n)}_\nu (t) ) / {\rm Det} ( \kappa^{(n)}_\nu (t^{}_{n+1}) )$ and $I^{}_{\rm Det} (t) =  {\rm Det} ({\kappa^{}_\nu} (t) ) / {\rm Det} ({\kappa^{}_\nu} (t^{}_4) )$, respectively, whereas $I^{\prime (n)}_{\rm Int}$ and $I^{}_{\rm int}$ are evaluated from gauge, Higgs and Yukawa couplings with the help of Eqs.~\eqref{eq:int-f1} and~\eqref{eq:int-f2}, respectively.}
	\label{fig:determinant}
\end{figure}

To illustrate the RG running behavior of the non-vanishing determinant against the energy scale $\mu$, we plot the natural logarithms of both $ I^{\prime (n)}_{\rm Det} (t)  = 	{\rm Det} ( \kappa^{(n)}_\nu (t) ) / {\rm Det} ( \kappa^{(n)}_\nu (t^{}_{n+1}) )$ and $I^{}_{\rm Det} (t) =  {\rm Det} ({\kappa^{}_\nu} (t) ) / {\rm Det} ({\kappa^{}_\nu} (t^{}_4) )$ together with those of $ I^{\prime (n)}_{\rm Int} \left(t,t^{}_{n+1}\right) $ and ${I^{}_{\rm Int}} \left(t \right) $ in Fig.~\ref{fig:determinant}. The former two are directly calculated from $\kappa^{}_\nu$ while the latter two are evaluated from the running gauge, Higgs and Yukawa couplings by means of Eqs.~\eqref{eq:int-f1} and~\eqref{eq:int-f2}, where the initial inputs used in Sec.~\ref{sec:eff-mass} are adopted. It shows the results obtained in two ways are consistent with each other, and the determinant is about 9/20 of its initial value at the GUT scale. 
\begin{figure}
	\centering
		\includegraphics[width=0.7\linewidth]{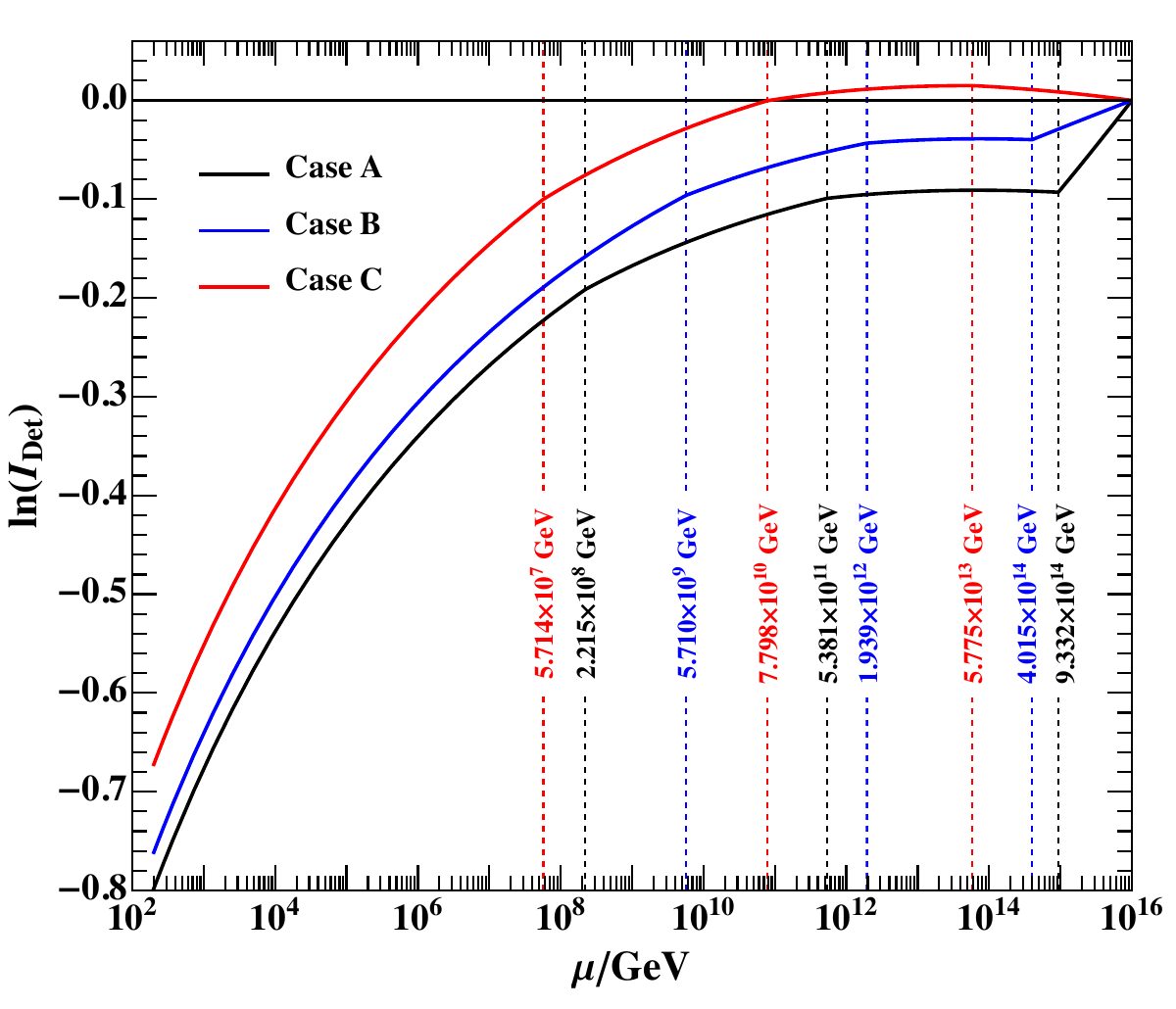}
	\vspace{-0.3cm}
	\caption{The natural logarithms of the ratio of the running determinant to its initial value against the energy scale, where $ I^{}_{\rm Det} (t) $ is directly calculated from $\kappa^{}_\nu$ via $I^{}_{\rm Det} (t) =  {\rm Det} ({\kappa^{}_\nu} (t) ) / {\rm Det} ({\kappa^{}_\nu} (t^{}_4) )$.}
	\label{fig:lnI3}
\end{figure}
Moreover, we take into account two other sets of inputs with different initial right-handed neutrino masses to show their effects on the RG running of the determinant. The natural logarithms of $I^{}_{\rm Det}$ against the energy scale with three different sets of initial inputs are shown in Fig.~\ref{fig:lnI3}, where the three cases are labeled as Case A, Case B, and Case C, respectively. The initial right-handed neutrino masses $\left( M^{}_1, M^{}_2, M^{}_3 \right)$ at $\mu =  1\times 10^{16}$ GeV in these three cases are
\begin{eqnarray}
		&&{\rm Case~A:~} \left( 2.215 \times 10^8,\; 5.382\times 10^{11},\; 9.620\times 10^{14} \right) {\rm GeV} \;,
		\nonumber
		\\
		&&{\rm Case~B:~} \left( 5.710 \times 10^9,\; 1.940 \times 10^{12},\; 4.082 \times 10^{14} \right) {\rm GeV} \;,
		\nonumber
		\\
		&&{\rm Case~C:~} \left( 5.714 \times 10^7,\; 7.799\times 10^{10},\; 5.796 \times 10^{13} \right) {\rm GeV} \;,
\end{eqnarray}
and the decoupling energy scales of right-handed neutrinos are explicitly shown in Fig.~\ref{fig:lnI3} for each case. Initial inputs for other couplings in these three cases are chosen to reproduce the same low-energy values of physical parameters. As shown in Fig.~\ref{fig:lnI3}, the determinants of effective neutrino mass matrix at the electroweak scale in three cases are all smaller than those at the GUT scale, but their running behaviors highly depend on right-handed neutrino masses and neutrino Yukawa couplings. In Case C, the neutrino Yukawa couplings are not large enough so that $3\alpha^{(3)}_1 + 2 {\rm Tr} \beta^{(3)}$ is negative above $\mu = M^{}_3$ and the determinant increases with the energy scale decreasing in this energy interval.

Turning to the singular values of $\kappa^{}_{\nu}$, we have
\begin{eqnarray}\label{eq:singular-det}
	\chi^{}_1 \left( t \right)  \chi^{}_2 \left( t \right)  \chi^{}_3 \left( t \right)  &=& I^{}_{\rm Int} \left( t \right) \cdot \chi^{}_1 \left( t^{}_4 \right)  \chi^{}_2 \left( t^{}_4 \right)  \chi^{}_3 \left( t^{}_4 \right) \;,\quad 
	\nonumber
	\\
	\varphi \left( t \right) &=& \varphi \left( t^{}_4 \right) \;,
\end{eqnarray}
from the absolute and argument parts of Eq.~\eqref{eq:detint}. The above results have exactly the same form as those in Eq.~\eqref{eq:singular-ab} but now contain threshold effects and are valid in the whole energy range. In addition, from ${\rm Tr} [  {\rm Adj} (\kappa^{(n)}_\nu ) C^{(n)}_5 ] = (3-n) {\rm Det} ( \kappa^{(n)}_\nu )$, one can acquire a correlation among the singular values $\chi^{(n)}_{i}$ and the diagonal elements of $\mathcal{C}^{(n)}_5 \equiv U^{(n)\dagger}_\nu C^{(n)}_5 U^{(n)\ast}_\nu $, namely
\begin{eqnarray}\label{eq:singular-all-2}
	(3-n)  \chi^{(n)}_1 \chi^{(n)}_2 \chi^{(n)}_3 &=&  \chi^{(n)}_2 \chi^{(n)}_3 ( \mathcal{C}^{(n)}_5 )^{}_{11} +  \chi^{(n)}_1  \chi^{(n)}_3 ( \mathcal{C}^{(n)}_5)^{}_{22} + \chi^{(n)}_1 \chi^{(n)}_2 ( \mathcal{C}^{(n)}_5)^{}_{33} \;.
\end{eqnarray}
If $ \chi^{(n)}_i$ (for $i=1,2,3$) are all non-zero, Eq.~\eqref{eq:singular-all-2} leads to $ \sum^{}_{i} {\rm Re} [ ( \mathcal{C}^{(n)}_5)^{}_{ii}/\chi^{(n)}_i ] = 3-n $ and $\sum^{}_{i} {\rm Im} [( \mathcal{C}^{(n)}_5)^{}_{ii}/\chi^{(n)}_i ] = 0$. These relations are invariant against the energy scale. 

On the other hand, starting with Eq.~\eqref{eq:kappanu}, one obtains
\begin{eqnarray}\label{eq:singular-all-1}
16 \pi^2 \frac{{\rm d} \chi^{(n)}_i} {{\rm d} t}  &=& \left(  \alpha^{(n)}_1 + 2 \beta^{\prime (n)}_{ii} - 2 \mathcal{T}^{(n)}_{ii} \right)  \chi^{(n)}_i  + \left( \alpha^{(n)}_2 -  \alpha^{(n)}_1 \right) ( \mathcal{C}^{(n)}_5)^{}_{ii} \;
\end{eqnarray}
with $ \mathcal{T}^{(n)} \equiv U^{(n)\dagger}_\nu {\rm d} U^{(n)}_\nu/ {\rm d} t$. It can be easily checked that Eq.~\eqref{eq:singular-all-1} produces the result in Eq.~\eqref{eq:det3} or ~\eqref{eq:singular-det} by making use of Eq.~\eqref{eq:singular-all-2}. Now, besides the terms proportional to $\chi^{(n)}_i$ itself, a term related to $ \mathcal{C}^{(n)}_5$ is involved. Consequently, the running of $\chi^{(n)}_i$ becomes more complicated compared with the case above the highest or below the lowest seesaw scale. Unlike those in Eq.~\eqref{eq:singular-ab-1}, it is difficult to get such a simple integral form for the separated singular values $\chi^{(n)}_i$ due to the entanglement among $\chi^{(n)}_i$ and $(\mathcal{C}^{(n)}_5)^{}_{ii}$. As a result, the RG running of $\chi^{(n)}_i$ among seesaw scales is not so analytically transparent as that above the highest or below the lowest seesaw scale. Moreover, the determinant only tells us the matrix rank is full or not and hence we do not know how the matrix rank behaves apart from keeping full or degenerate. This means it is still unclear if any other neutrino may become massless during the RG running when one of three neutrinos is initially massless. One has to include other two matrix invariants, such as $I^{}_1 = {\rm Tr} H^{}_\nu$ and $ I^{}_2 =  [({\rm Tr} H^{}_\nu )^2 - {\rm Tr} (H^2_\nu) ]/2$ with $H^{}_\nu = \kappa^{}_\nu \kappa^\dagger_\nu$, to fully capture the matrix rank. It is quite challenging to analytically figure out whether $I^{}_{1,2}$ can become vanishing if they are initially non-zero (but it expects not). Nevertheless, the conclusion on the massless neutrino is still valid no matter how $I^{}_{1,2}$ behave against the energy scale.

\section{Summary}

An essential question in neutrino physics concerns the determination of absolute neutrino masses, specifically that of the lightest neutrino. Current terrestrial experiments and cosmological observations still allow the lightest neutrino to be massless. However, the vanishing neutrino mass is not protected by any fundamental symmetry generally and hence the lightest neutrino may acquire a non-zero mass induced by quantum corrections even it is initially massless. In this work, we provide a strict proof that the vanishing neutrino mass is stable against one-loop RG corrections including threshold effects in the (minimal) type-I seesaw mechanism. To achieve this conclusion, we derive the complete one-loop RGEs of all dim-5 operators in the $\nu$SMEFT and then apply them to the one-loop RGEs among hierarchy seesaw scales in the type-I seesaw mechanism. We find some new contributions to the RGEs of $\lambda$, $Y^{}_\nu$ and $C^{}_5$ that have been overlooked for quite a long time. Based on the corrected RGEs, we derive the equation for the determinant of the effective neutrino mass matrix, that governs the RG running behavior of the determinant against the energy scale. It shows that the determinant is proportional to its initial value and hence it remains vanishing during RG running if it initially vanishes. This leads us to the aforementioned conclusion about the massless neutrino.

It is worth pointing out that conclusions in the cases with more than three right-handed neutrinos are still unclear. However, such cases are not well motivated for three generations of active neutrinos and not taken into consideration in this work. The other concern is that the determinant itself can not fully determine the matrix rank. One has to introduce other matrix invariants (e.g., $I^{}_{1,2}$) to completely determine the rank. But this issue does not affect the conclusion we have achieved and we leave discussions on this issue somewhere else.

\begin{acknowledgments}
We are very grateful to Prof. Alejandro Ibarra for useful discussions and to Prof. Zhi-zhong Xing for carefully reading this manuscript. This work is supported by the Alexander von Humboldt Foundation. We would also like to express special thanks both to the Southampton High Energy Physics group at the School of Physics and Astronomy, University of Southampton and to the Mainz Institute for Theoretical Physics (MITP) of the Cluster of Excellence PRISMA+ (Project ID 390831469), for their hospitality and support.
\end{acknowledgments}

\appendix

\section{\label{app:A}Details for New Contributions to RGEs between Seesaw Scales}

Here we provide some details for the overlooked contributions to the RGEs of the neutrino Yukawa coupling matrix $Y^{}_\nu$, the Higgs quartic coupling $\lambda$, and the Wilson coefficient $C^{}_5$ of the Weinberg operator between seesaw scales in the type-I seesaw mechanism that are governed by diagrams shown in Fig.~\ref{fig:three-point}. We start with the amplitude for Diagram (a) in Fig.~\ref{fig:three-point}, namely~\footnote{We adopt the Feynman rules for Majorana particles proposed in~\cite{Denner:1992vza,Denner:1992me}.}
\begin{eqnarray}
	\rmi \mathcal{M}^{}_a &=& \int \frac{{\rm d}^d k}{\left( 2\pi\right)^d} \overline{u}^{}_\ell \left( p^\prime \right) \rmi C^{\alpha\gamma}_5 \left( \epsilon^{ad}\epsilon^{bc} + \epsilon^{ac}\epsilon^{bd} \right) P^{}_{\rm R} \frac{\rmi}{-\slashed{k}} \left( -\rmi \right) \epsilon^{bd} \left( Y^\ast_\nu \right)^{}_{\gamma\beta} P^{}_{\rm L} u^{}_N \left( p \right) \frac{\rmi}{\left( k+p \right)^2 - m^2}
	\nonumber
	\\
	&=& 3\epsilon^{ac} \left( C^{}_5 Y^\ast_\nu \right)^{}_{\alpha\beta} \overline{u}^{}_\ell \left( p^\prime \right) \left\{ \int \frac{{\rm d}^d k}{\left( 2\pi\right)^d} \frac{\slashed{k}}{k^2 \left[ \left( k+p\right)^2 - m^2 \right] }  \right\} P^{}_{\rm L} u^{}_N \left( p \right) 
	\nonumber
	\\
	&=& - \frac{\rmi}{ 16\pi^2 \varepsilon} \frac{3}{2} \epsilon^{ac} \left( C^{}_5 Y^\ast_\nu \right)^{}_{\alpha\beta}  \overline{u}^{}_\ell \left( p^\prime \right) \slashed{p} u^{}_N \left( p \right) + {\rm UV ~finite} \;.
\end{eqnarray}
It is obvious that the divergence in the above equation can not be eliminated by the neutrino Yukawa coupling counterterm. Its root cause is that Diagram (a) breaks the lepton number by 2 units while the Yukawa coupling interaction does not. To cancel out this divergence, the operator $\Opr^{}_{\ell HN1}$ is indispensable and its counterterm turns out to be
\begin{eqnarray}\label{eq:lhn1}
	\delta G^{}_{\ell HN1} &=&  \frac{3 \rmi C^{}_5 Y^\ast_\nu}{32\pi^2 \varepsilon} \;.
\end{eqnarray}
This contribution can be converted into those of $Y^{}_\nu$ and $C^{}_5$ by means of the right-handed neutrinos' EoM, that is
\begin{eqnarray}\label{eq:lhn1-eom}
	\delta G^{\alpha\beta}_{\ell HN1} \Opr^{\alpha\beta}_{\ell HN1} &=& -\rmi \left( \delta G^{}_{\ell HN1} M^{}_N \right)^{}_{\alpha\beta} \overline{\ell^{}_{\alpha\rm L}} \widetilde{H} N^{}_{\beta\rm R} - \rmi \left( \delta G^{}_{\ell HN1} Y^{\rm T}_\nu \right)^{}_{\alpha\beta} \overline{\ell^{}_{\alpha\rm L}} \widetilde{H} \widetilde{H}^{\rm T} \ell ^{\rm c}_{\beta \rm L} \;.
\end{eqnarray}
With the help of Eqs.~\eqref{eq:lhn1} and~\eqref{eq:lhn1-eom}, one has
\begin{eqnarray}
	\delta Z^{\rm (add)}_{Y_\nu} &=&  \rmi Y^{-1}_\nu \delta G^{}_{\ell HN1} M^{}_N = - \loopf \frac{3}{2} Y^{-1}_\nu C^{}_5 Y^\ast_\nu M^{}_N \;,
	\nonumber
	\\
	\delta C^{\rm (add)}_5 &=& -  \rmi \left( \delta G^{}_{\ell HN1} Y^{\rm T}_\nu + Y^{}_\nu \delta G^{\rm T}_{\ell HN1} \right) = \loopf \frac{3}{2} \left[ Y^{}_\nu Y^\dagger_\nu C^{}_5 + C^{}_5 \left( Y^{}_\nu Y^\dagger_\nu \right)^{\rm T} \right] \;,
\end{eqnarray}
leading to the additional contributions to the RGEs of $Y^{}_\nu$ and $C^{}_5$ in Eq.~\eqref{eq:type-i}.

For Diagram (b) in Fig.~\ref{fig:three-point}, the corresponding amplitude (including the one with the $\mathcal{O}^{}_5$ vertex and $N^{}_i$ propagator exchanged) is found to be 
\begin{eqnarray}\label{eq:ampb}
	\rmi \mathcal{M}^{}_{b} &=& - \int \frac{{\rm d}^d k}{\left( 2\pi\right)^d} {\rm Tr} \left[ \rmi C^{\alpha\beta}_5 P^{}_{\rm R} \left( \epsilon^{ec}\epsilon^{fd} + \epsilon^{ed} \epsilon^{fc} \right) \frac{\rmi}{\slashed{k}}\left( -\rmi \right) \left( Y^\ast_\nu \right)^{}_{\beta i} P^{}_{\rm L} \epsilon^{ea} \frac{\rmi}{\slashed{k}-M^{}_i} \left( -\rmi \right) \left( Y^\ast_\nu \right)^{}_{\alpha i} P^{}_{\rm L} \epsilon^{fb} \frac{\rmi}{\slashed{k}} \right.
	\nonumber
	\\
	&& + \left.  \rmi C^{\ast\alpha\beta}_5 P^{}_{\rm L} \left( \epsilon^{ea}\epsilon^{fb} + \epsilon^{eb} \epsilon^{fa} \right) \frac{\rmi}{\slashed{k}}\left( -\rmi \right) \left( Y^{}_\nu \right)^{}_{\beta i} P^{}_{\rm R} \epsilon^{ec} \frac{\rmi}{\slashed{k}-M^{}_i} \left( -\rmi \right) \left( Y^{}_\nu \right)^{}_{\alpha i} P^{}_{\rm R} \epsilon^{fd} \frac{\rmi}{\slashed{k}} \right]
	\nonumber
	\\
	&=& 2 \left( \delta^{ac}\delta^{bd} + \delta^{ad} \delta^{bc} \right) \left[ \left( C^{}_5 Y^\ast_\nu \right)^{}_{\alpha i} M^{}_i \left( Y^\dagger_\nu \right)^{}_{i \alpha} + \left( Y^{}_\nu \right)^{}_{\alpha i} M^{}_i \left( Y^{\rm T}_\nu C^\dagger_5  \right)^{}_{i\alpha} \right] \int \frac{{\rm d}^d k}{\left( 2\pi\right)^d}  \frac{1}{k^2 \left( k^2 - M^2_i  \right)}
	\nonumber
\\
&=& \frac{2\rmi}{16\pi^2\varepsilon}  \left( \delta^{ac}\delta^{bd} + \delta^{ad} \delta^{bc} \right) {\rm Tr} \left( C^{}_5 Y^\ast_\nu M^{}_N Y^\dagger_\nu + Y^{}_\nu M^\dagger_N Y^{\rm T}_\nu C^\dagger_5  \right) + {\rm UV ~finite} \;.
\end{eqnarray}
In the above amplitude, we have taken all external momentum to be zero since we focus on contributions to $\lambda$ and four Higgs external lines are already mass-dimension four. The UV divergent part in Eq.~\eqref{eq:ampb} leads to an additional contribution to the renormalization constant $\delta Z^{}_\lambda$, i.e.,
\begin{eqnarray}
	\delta Z^{\rm (add)}_\lambda \lambda = \frac{1}{16\pi^2\varepsilon}  {\rm Tr} \left( C^{}_5 Y^\ast_\nu M^{}_N Y^\dagger_\nu + Y^{}_\nu M^\dagger_N Y^{\rm T}_\nu C^\dagger_5  \right) \;,
\end{eqnarray}
which results in new terms involving $C^{}_5$ in the RGE of $\lambda$ as shown in Eq.~\eqref{eq:type-i}.

\section{\label{app:adj} The Derivation of Eqs.~\eqref{eq:dett1} and \eqref{eq:dett2}}
For a $3\times3$ matrix $M$, its adjugate can be written as
\begin{eqnarray}
	{\rm Adj} (M) &=& \frac{1}{2}  \left[ \left( {\rm Tr} (M) \right)^2 - {\rm Tr} \left( M^2\right) \right] \mathbf{1} - M {\rm Tr}(M) + M^2 \;,
	\nonumber
	\\
\end{eqnarray}
from which, one can get
\begin{eqnarray}\label{eq:adj1}
	{\rm Adj} (M+W) &=& {\rm Adj} (M) + {\rm Adj} (W)  - M {\rm Tr} (W) - W {\rm Tr} (M)  + MW + WM 
	\nonumber
	\\
	&& + \left[  {\rm Tr} (M) {\rm Tr} (W) - {\rm Tr} (MW) \right] \mathbf{1} \;.\;\;\;
\end{eqnarray}
With the help of Eq.~\eqref{eq:adj1}, one achieves
\begin{eqnarray}\label{eq:adj2}
	{\rm Tr} \left( {\rm Adj} (M+W) W \right) &=& 3 {\rm Det} (W) + 2 {\rm Tr} \left( {\rm Adj} (W) M  \right)  + {\rm Tr} \left( {\rm Adj} (M) W\right) \;
\end{eqnarray}
with $M$ and $W$ are both $3\times3$ matrices. Substituting $(\kappa^{(n)}_D, C^{(n)}_5)$ for $(M,W)$ or $(W,M)$ in Eq.~\eqref{eq:adj2}, we have
\begin{eqnarray}\label{eq:adj3}
	{\rm Tr} \left( {\rm Adj} (\kappa^{(n)}_\nu) C^{(n)}_5 \right) &=& 3 {\rm Det} (C^{(n)}_5) + 2 {\rm Tr} \left( {\rm Adj} (C^{(n)}_5) \kappa^{(n)}_D  \right)  + {\rm Tr} \left( {\rm Adj} (\kappa^{(n)}_D) C^{(n)}_5\right)  \;,
	\nonumber
	\\
	{\rm Tr} \left( {\rm Adj} (\kappa^{(n)}_\nu) \kappa^{(n)}_D \right) &=& 3 {\rm Det} (\kappa^{(n)}_D) + 2 {\rm Tr} \left( {\rm Adj} (\kappa^{(n)}_D) C^{(n)}_5  \right)  + {\rm Tr} \left( {\rm Adj} (C^{(n)}_5) \kappa^{(n)}_D  \right)  \;.
\end{eqnarray}
The above two equations in Eq.~\eqref{eq:adj3} are independent of the energy scale and the combination of them leads us to 
\begin{eqnarray}\label{eq:adj4}
	{\rm Det} (\kappa^{(n)}_\nu) &=& {\rm Det} (C^{(n)}_5) + {\rm Det} (\kappa^{(n)}_D) + {\rm Tr} \left( {\rm Adj} (C^{(n)}_5) \kappa^{(n)}_D  \right) + {\rm Tr} \left( {\rm Adj} (\kappa^{(n)}_D) C^{(n)}_5\right) \;.
\end{eqnarray}
Now taking advantage of Eqs.~\eqref{eq:adj3} and \eqref{eq:adj4}, the energy-scale-independent relations given by Eqs.~\eqref{eq:dett1} and \eqref{eq:dett2} can be easily derived.

\bibliographystyle{JHEP}
\bibliography{threshold_effects}

\end{document}